\documentclass[journal]{IEEEtran}
\ifCLASSINFOpdf
\else
\fi

\input epsf
\usepackage{graphicx}
\usepackage{subfig}
\usepackage{xcolor}
\usepackage{amsmath}
\usepackage{epstopdf}
\usepackage{cite}
\begin{document}
%
% paper title
% Titles are generally capitalized except for words such as a, an, and, as,
% at, but, by, for, in, nor, of, on, or, the, to and up, which are usually
% not capitalized unless they are the first or last word of the title.
% Linebreaks \\ can be used within to get better formatting as desired.
% Do not put math or special symbols in the title.
\title{\LARGE Pencil-Beam Single-point-fed Dirac Leaky-Wave Antenna\\on a Transmission-Line Grid }
%
%
% author names and IEEE memberships
% note positions of commas and nonbreaking spaces ( ~ ) LaTeX will not break
% a structure at a ~ so this keeps an author's name from being broken across
% two lines.
% use \thanks{} to gain access to the first footnote area
% a separate \thanks must be used for each paragraph as LaTeX2e's \thanks
% was not built to handle multiple paragraphs
%

\author{Ayman~H.~Dorrah,~\IEEEmembership{Student Member,~IEEE}~and~George~V.~Eleftheriades,~\IEEEmembership{Fellow,~IEEE}% <-this % stops a space
\thanks{The authors are with the Edward S. Rogers, Sr. Department of Electrical and	Computer Engineering, University of Toronto, Toronto, ON M5S 3G4, Canada	(e-mail: ayman.dorrah@mail.utoronto.ca).}}

% note the % following the last \IEEEmembership and also \thanks - 
% these prevent an unwanted space from occurring between the last author name
% and the end of the author line. i.e., if you had this:
% 
% \author{....lastname \thanks{...} \thanks{...} }
%                     ^------------^------------^----Do not want these spaces!
%
% a space would be appended to the last name and could cause every name on that
% line to be shifted left slightly. This is one of those "LaTeX things". For
% instance, "\textbf{A} \textbf{B}" will typeset as "A B" not "AB". To get
% "AB" then you have to do: "\textbf{A}\textbf{B}"
% \thanks is no different in this regard, so shield the last } of each \thanks
% that ends a line with a % and do not let a space in before the next \thanks.
% Spaces after \IEEEmembership other than the last one are OK (and needed) as
% you are supposed to have spaces between the names. For what it is worth,
% this is a minor point as most people would not even notice if the said evil
% space somehow managed to creep in.

% The paper headers
\markboth{IEEE Antennas and Wireless Propagation Letters,~Vol.~**, No.~*, ******~2016}%
{}
% The only time the second header will appear is for the odd numbered pages
% after the title page when using the twoside option.
% 
% *** Note that you probably will NOT want to include the author's ***
% *** name in the headers of peer review papers.                   ***
% You can use \ifCLASSOPTIONpeerreview for conditional compilation here if
% you desire.

% If you want to put a publisher's ID mark on the page you can do it like
% this:
%\IEEEpubid{0000--0000/00\$00.00~\copyright~2015 IEEE}
% Remember, if you use this you must call \IEEEpubidadjcol in the second
% column for its text to clear the IEEEpubid mark.

% use for special paper notices
%\IEEEspecialpapernotice{(Invited Paper)}

% make the title area
\maketitle

% As a general rule, do not put math, special symbols or citations
% in the abstract or keywords.
\begin{abstract}
Leaky-wave antennas (LWAs) are widely used as single-point-fed linear antenna arrays. The extension of LWAs to 2D implies that they can be used as single-point-fed 2D antenna arrays without requiring a complex feeding network. However, generating a pencil beam from 2D LWAs is not straightforward and due care has to be taken for the design of the LWA. On the other hand, transmission-line (TL) grids have demonstrated interesting behaviors, such as an effective negative refractive index and growing of evanescent waves. In this paper, a single-point-fed TL-grid 2D Dirac leaky-wave antenna (DLWA) design is proposed that generates a pencil beam at both broadside and slightly tilted angles. The TL-grid unit cell is analytically treated in light of its scattering and impedance matrices. The optimized TL-grid unit cell is shown to exhibit a closed bandgap in the dispersion relation which is also linearly varying with frequency (hence it is a DLWA). The proposed 2D DLWA design is fabricated and the experimental results are presented.
\end{abstract}

% Note that keywords are not normally used for peerreview papers.

% For peer review papers, you can put extra information on the cover
% page as needed:
% \ifCLASSOPTIONpeerreview
% \begin{center} \bfseries EDICS Category: 3-BBND \end{center}
% \fi
%
% For peerreview papers, this IEEEtran command inserts a page break and
% creates the second title. It will be ignored for other modes.
\IEEEpeerreviewmaketitle

\section{Introduction}
There is ever increasing demand for low profile antennas with highly directive pencil beams especially for satellite communications. Two-dimensional (2D) antenna arrays are typically used for this purpose because of their low profile and capability to produce highly directive pencil beams with both azimuth and elevation beam-scanning capabilities. However, in traditional designs, 2D antenna arrays require complex feeding networks to excite each element of the antenna array with the correct phase and power level. Another challenge in 2D arrays is the inherent losses in the complex feeding networks especially at high frequencies. 

To eliminate the feeding network entirely, there has been a lot of research interest in using metamaterials to design 2D leaky-wave antenna (LWA) arrays which typically have high gain and low profile. A single-point-fed 2D LWA design was proposed and experimentally demonstrated in \cite{CalozConical}. The antenna is center-fed and due to the azimuthal symmetry, it produces a conical beam with elevation angle that is scanned by sweeping the input frequency. However, the produced conical beam is not suitable for applications that require high directivity. On the other hand, holographic techniques have been proposed to generate a pencil beam from single-point-fed 2D holographic LWAs \cite{SievenpiperEllipse,MaciAll}. A surface wave must be excited in order for these designs to work properly and achieve the required interference pattern. Furthermore, the proposed holographic designs do not achieve a true broadside operation and are not meant for beam scanning; moreover, they are difficult to design.

From the previous discussion, it is evident that the single-point-fed 2D LWAs proposed in the literature generate a conical beam or only generate a pencil beam at a fixed radiation direction. Thus, a single-point-fed 2D LWA design that operates in transmission-line (TL) form, capable of generating a directive scannable pencil beam would be a desirable addition to the state of the art. In this paper, we report the design, fabrication and experimental testing of a low-profile corner-fed 2D LWA that is based on TL-grids capable of generating a highly directive pencil beam with continuous scanning through broadside. It is shown that the dispersion relation of the optimized TL-grid unit cell is linearly varying with frequency with a closed bandgap at the $\Gamma$-point. This means that the TL-grid unit cell exhibits a Dirac dispersion and the resulting 2D antenna design is a Dirac leaky-wave antenna (DLWA) analogous to the photonic crystal antenna proposed in \cite{Moh_LWA}. The antenna is operated in its dominant mode, thus, it can be fed by a transmission line allowing for relatively wideband operation and a decent scanning range with frequency.   

\section{Analysis of the TL-grid Unit cell}

The model of an ideal TL-grid unit cell terminated with $Z_o$ is presented in Fig.~\ref{fig:UnitCell_Model_Ideal_TLGrid_Single}. Using transmission-line theory \cite{Pozar}, the S-parameters of the TL-grid unit cell can be analytically obtained as: 
\begin{figure}  
	\centering
	\includegraphics[width=0.48\textwidth,trim={{0.06\textwidth} {0.08\textwidth} {0.05\textwidth} {0.05\textwidth}},clip]{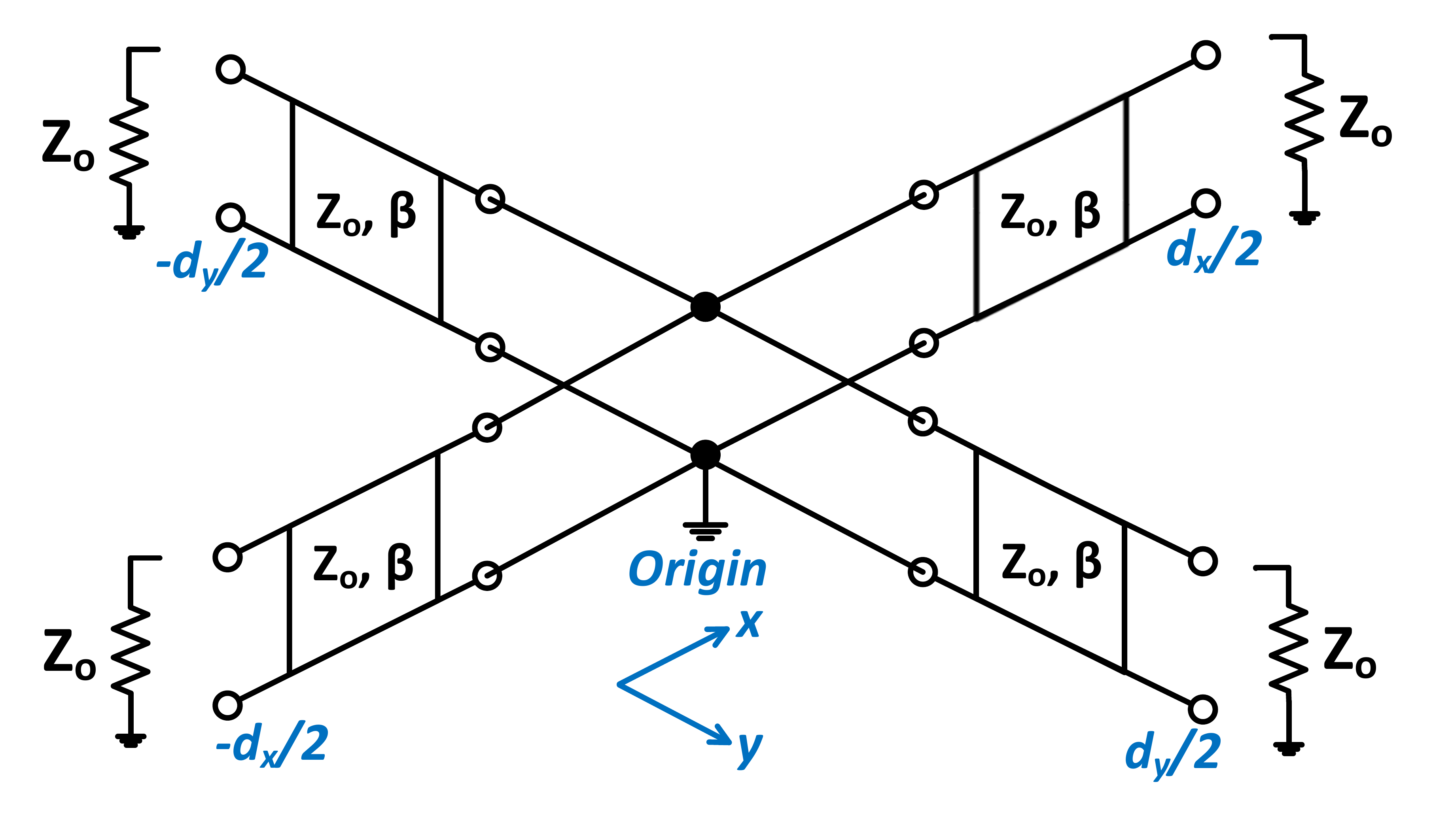}
	\caption{The unit cell model of an ideal 2D TL-grid unit cell terminated with $Z_o$.}
	\label{fig:UnitCell_Model_Ideal_TLGrid_Single}  
\end{figure}
\begin{equation}\label{eq:S_matrix}
\boldsymbol{S}=\frac{1}{2}\chi\psi\left[ 
\begin{array}{cccc}
-\chi/\psi& 1& \chi/\psi& 1\\
1& -\psi/\chi& 1& \psi/\chi\\
\chi/\psi& 1& -\chi/\psi& 1\\
1& \psi/\chi& 1&-\psi/\chi 							 							 
\end{array} 
\right]
\end{equation}
where $\chi = e^{-j\beta d_x/2}$ and $\psi = e^{-j\beta d_y/2}$, $\beta$ is the propagation constant of the transmission lines used, $d_x$ and $d_y$ are the dimensions of the TL-grid unit cell along the $x$ and $y$ directions, respectively. Assuming a symmetric TL-grid unit cell with $d_x=d_y=d$, the S-matrix in (\ref{eq:S_matrix}) can be simplified to:
\begin{equation}\label{eq:S_matrix1X1}
\boldsymbol{S_{d\times d}}=\frac{1}{2}e^{-j\beta d}\left[ 
\begin{array}{cccc}
-1& 1& 1& 1\\
1& -1& 1& 1\\
1& 1& -1& 1\\
1& 1& 1&-1
\end{array} 
\right]
\end{equation}

\begin{figure}
	\centering{
		\subfloat{\includegraphics[width=0.24\textwidth,trim={{0.02\textwidth} {0} {0.08\textwidth} {0.03\textwidth}},clip]{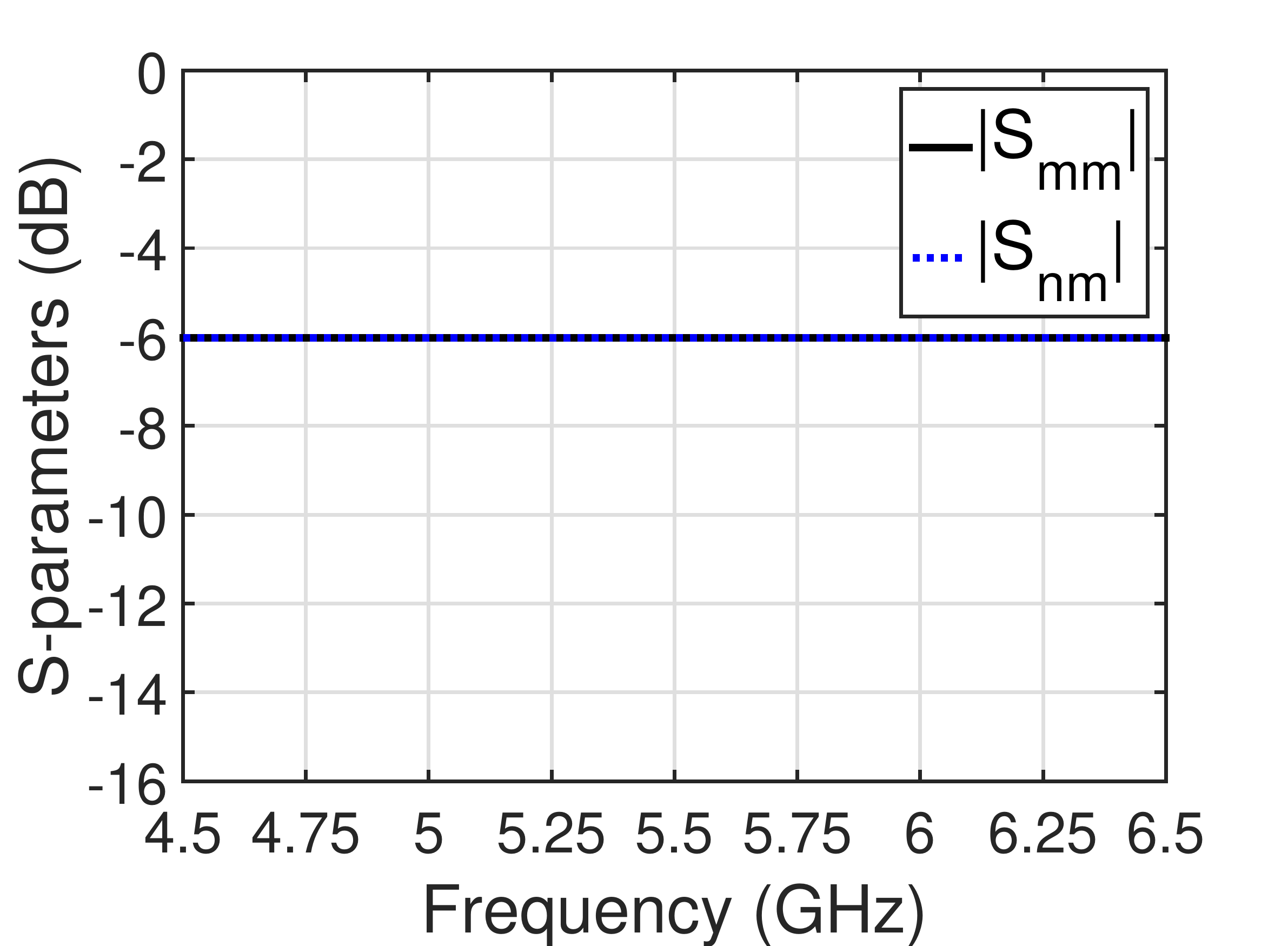}}
		\subfloat{\includegraphics[width=0.24\textwidth,trim={{0.02\textwidth} {0} {0.08\textwidth} {0.03\textwidth}},clip]{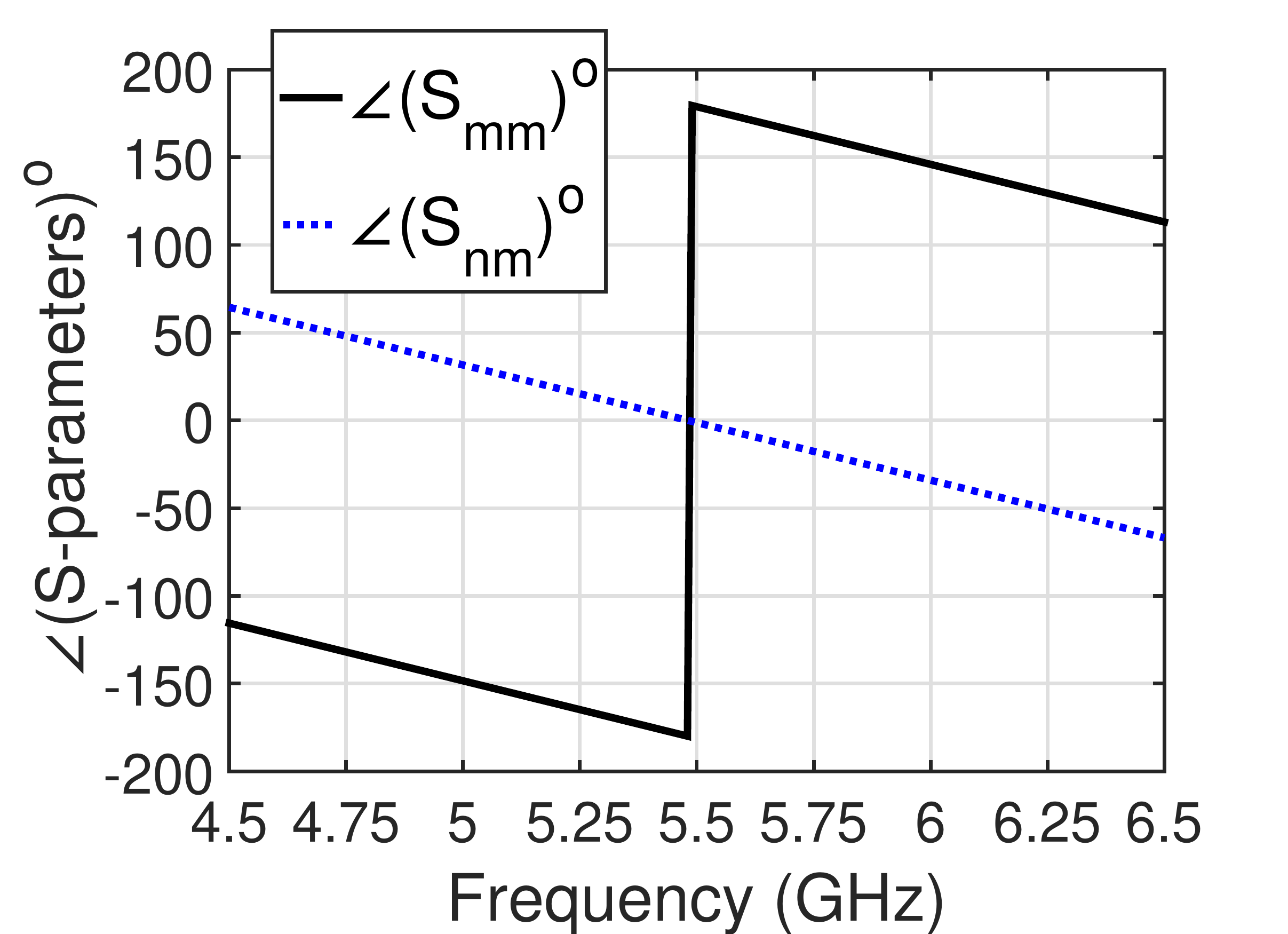}}
	}
	\caption{Ideal S-parameter response of the symmetric TL-grid unit cell with $d=40mm$ and $\beta d=2\pi$ at $5.48GHz$.}
	\label{fig:SPar_SingleUnitCell_Model_Ideal_TLGrid}
\end{figure}

The magnitude and phase of $S_{mm}$ and $S_{nm}$ ($m\neq n$) for a symmetric TL-grid unit cell are plotted in Fig.~\ref{fig:SPar_SingleUnitCell_Model_Ideal_TLGrid}. It is clear that the magnitude of all the scattering parameters is equal to $6dB$ regardless of the dimensions of the TL-grid unit cell, which means that an incident voltage wave at one of the ports is distributed equally between the $4$ ports of the TL-grid unit cell. In addition, the voltage waves leaving the ports of the TL-grid unit cell are in phase with the incident voltage wave at the frequency where $\beta d=2\pi$. This frequency will be referred to as the $\Gamma$-point frequency because at that frequency, an infinitely periodic TL-grid structure has a single operation point in the 2D dispersion relation which is at the $\Gamma$-point \cite{Andrade}. 

The impedance matrix of the symmetric TL-grid unit cell can be derived from the S-matrix in (\ref{eq:S_matrix1X1}) \cite{Pozar}: 
\begin{equation}
\boldsymbol{Z_{d\times d}}=Z_o\frac{e^{j\beta d}}{e^{2j\beta d}-1}\left[ 
\begin{array}{cccc}
\delta & 1& 1& 1\\
1& \delta & 1& 1\\
1& 1& \delta & 1\\
1& 1& 1& \delta
\end{array} 
\right]
\end{equation}
where $ \delta = 2cos(\beta d)-1$.

\begin{figure}
	\centerline{
		\subfloat{
			\includegraphics[width=0.24\textwidth,trim={{0.02\textwidth} {0} {0.08\textwidth} {0.03\textwidth}},clip]{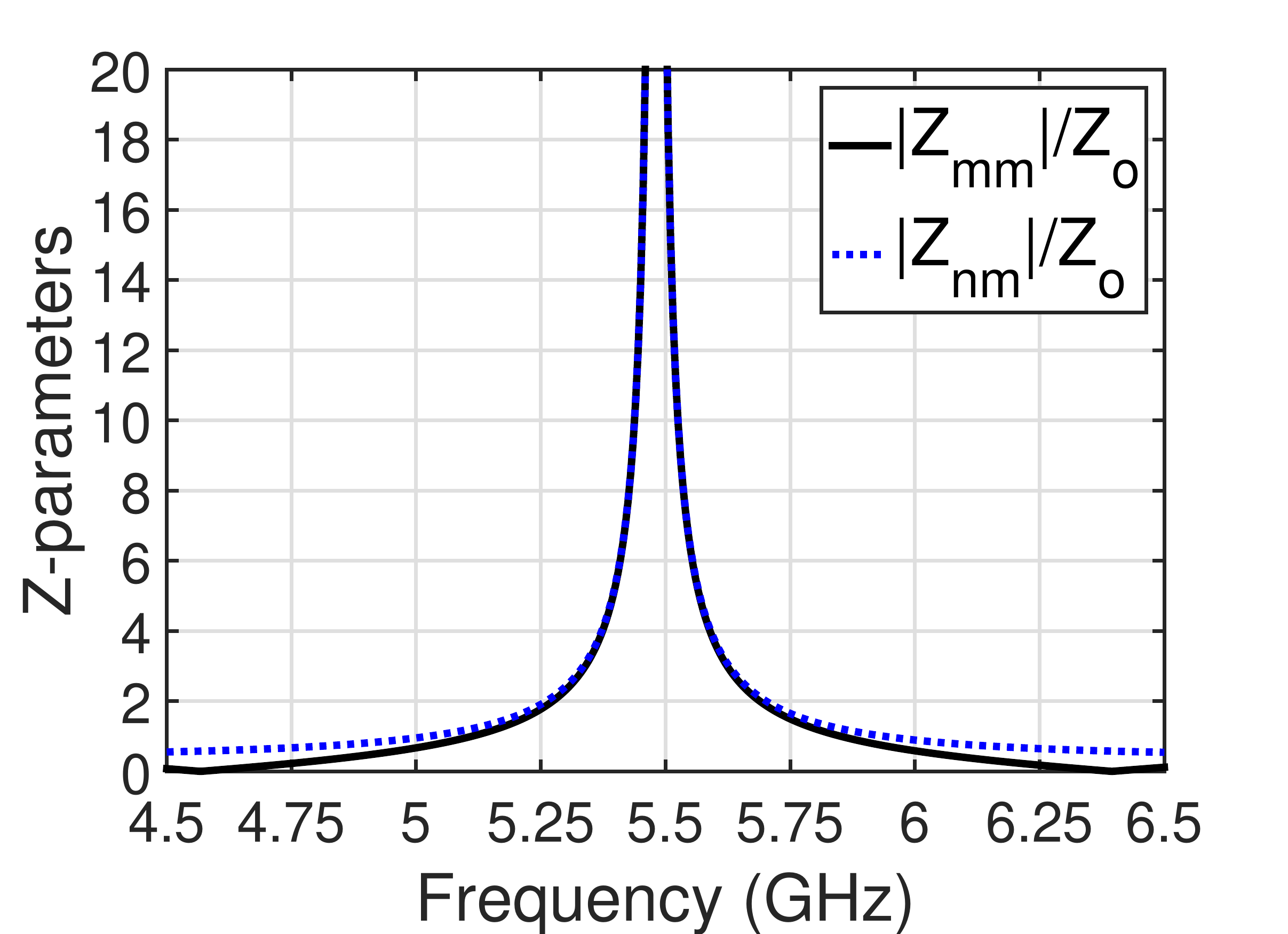}}
		\subfloat{
			\includegraphics[width=0.24\textwidth,trim={{0.02\textwidth} {0} {0.08\textwidth} {0.03\textwidth}},clip]{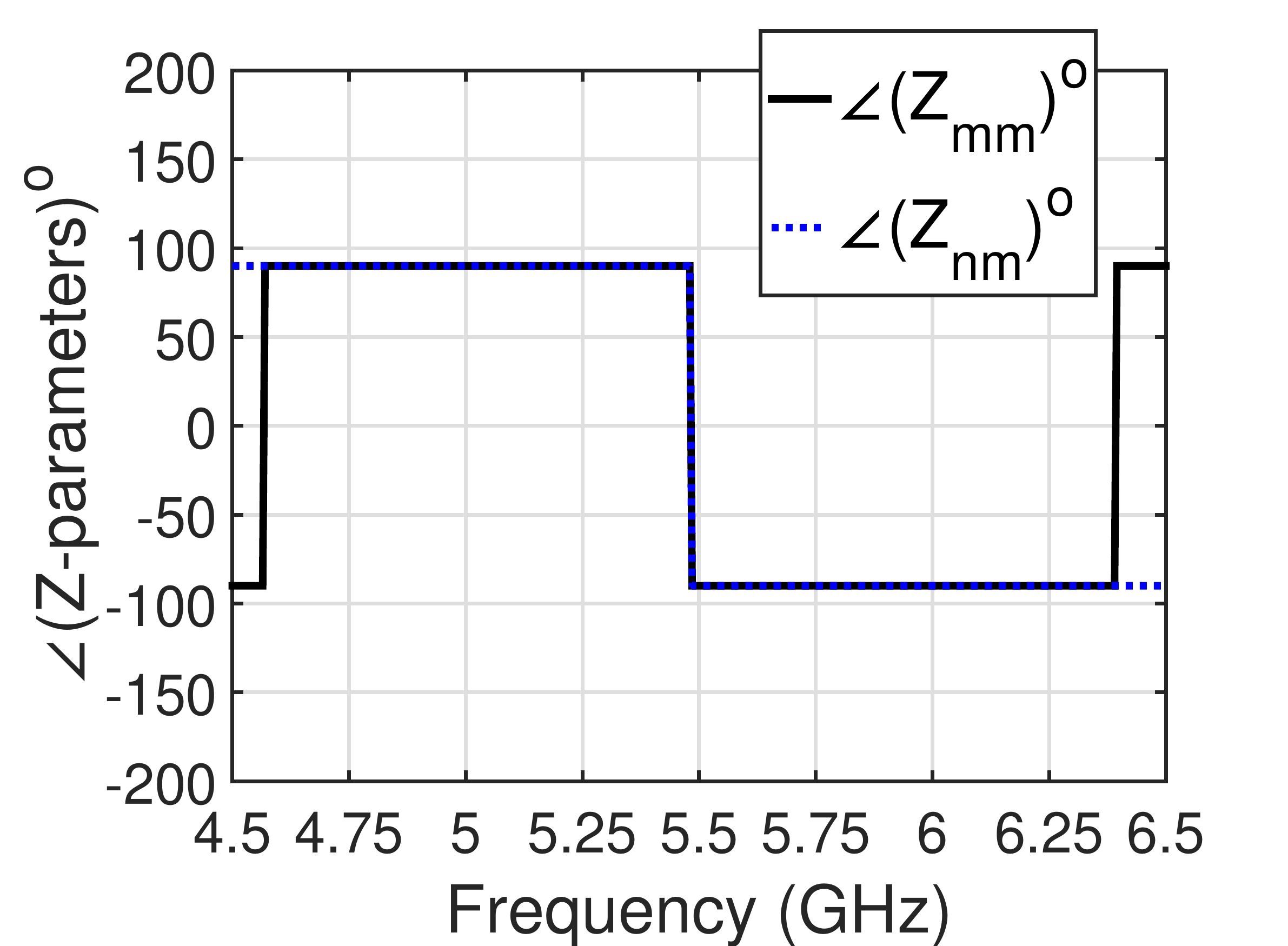}}}
	\caption{Ideal Z-parameter response of a symmetric TL-grid unit cell with $d=40mm$ and $\beta d=2\pi$ at $5.48GHz$.}
	\label{fig:ZPar_SingleUnitCell_Model_Ideal_TLGrid}
\end{figure}

The magnitude and phase plots of the Z-parameters of the symmetric TL-grid unit cell are presented in Fig.~\ref{fig:ZPar_SingleUnitCell_Model_Ideal_TLGrid}. At the $\Gamma$-point frequency, the Z-matrix elements are infinite in magnitude. This singularity is associated with the vanishing denominator ($e^{2j\beta d}-1$) when $d=2\pi/\beta$. 

\section{2D Single-point Fed TL-grid Array Structures}

\begin{figure}
	\centerline{
		\subfloat{
			\includegraphics[width=0.24\textwidth]{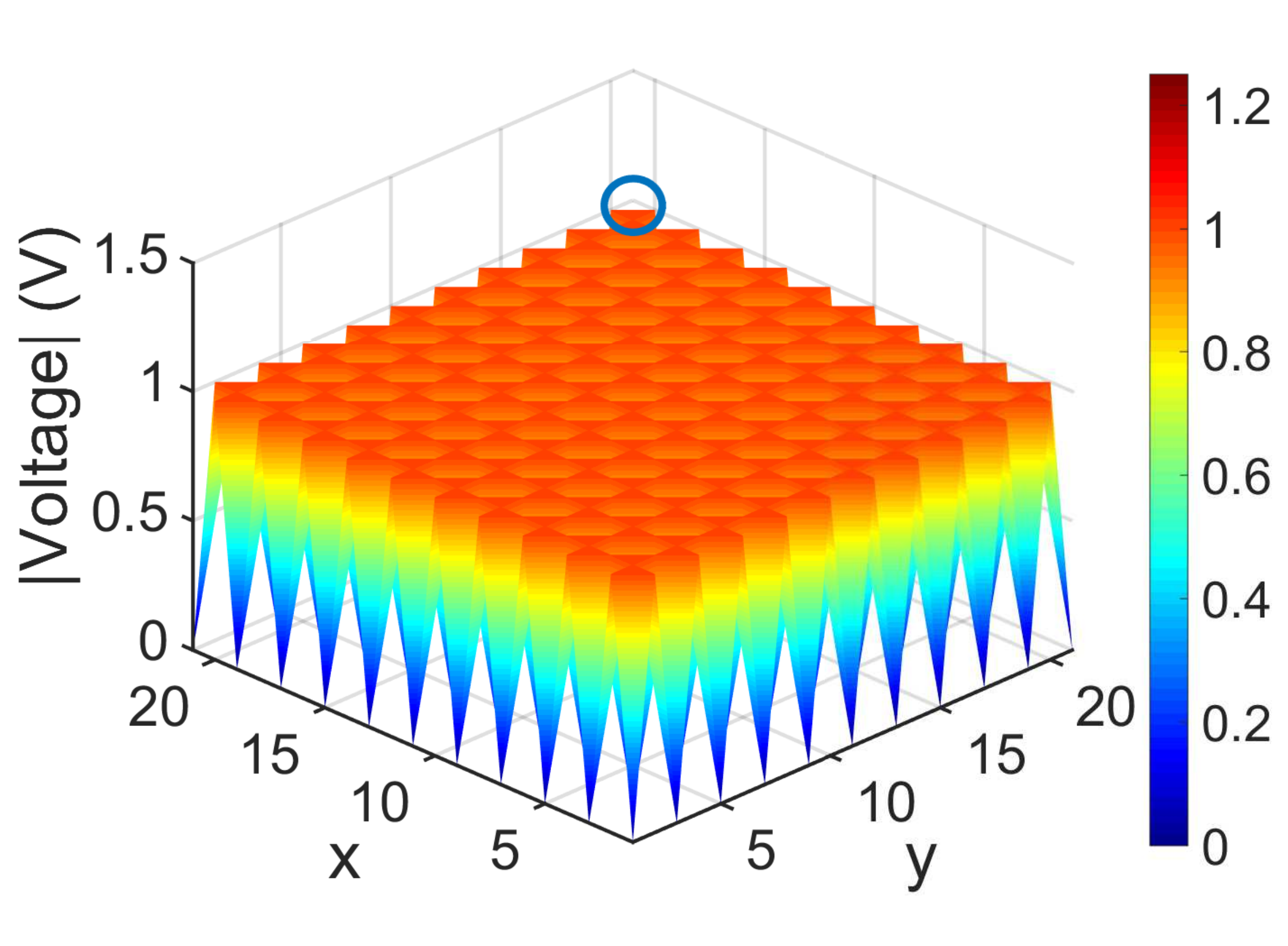}
			\label{fig:VoltageDist_10x10_Ideal_TLGrid_Gamma_Dirac}}
		\subfloat{
			\includegraphics[width=0.24\textwidth]{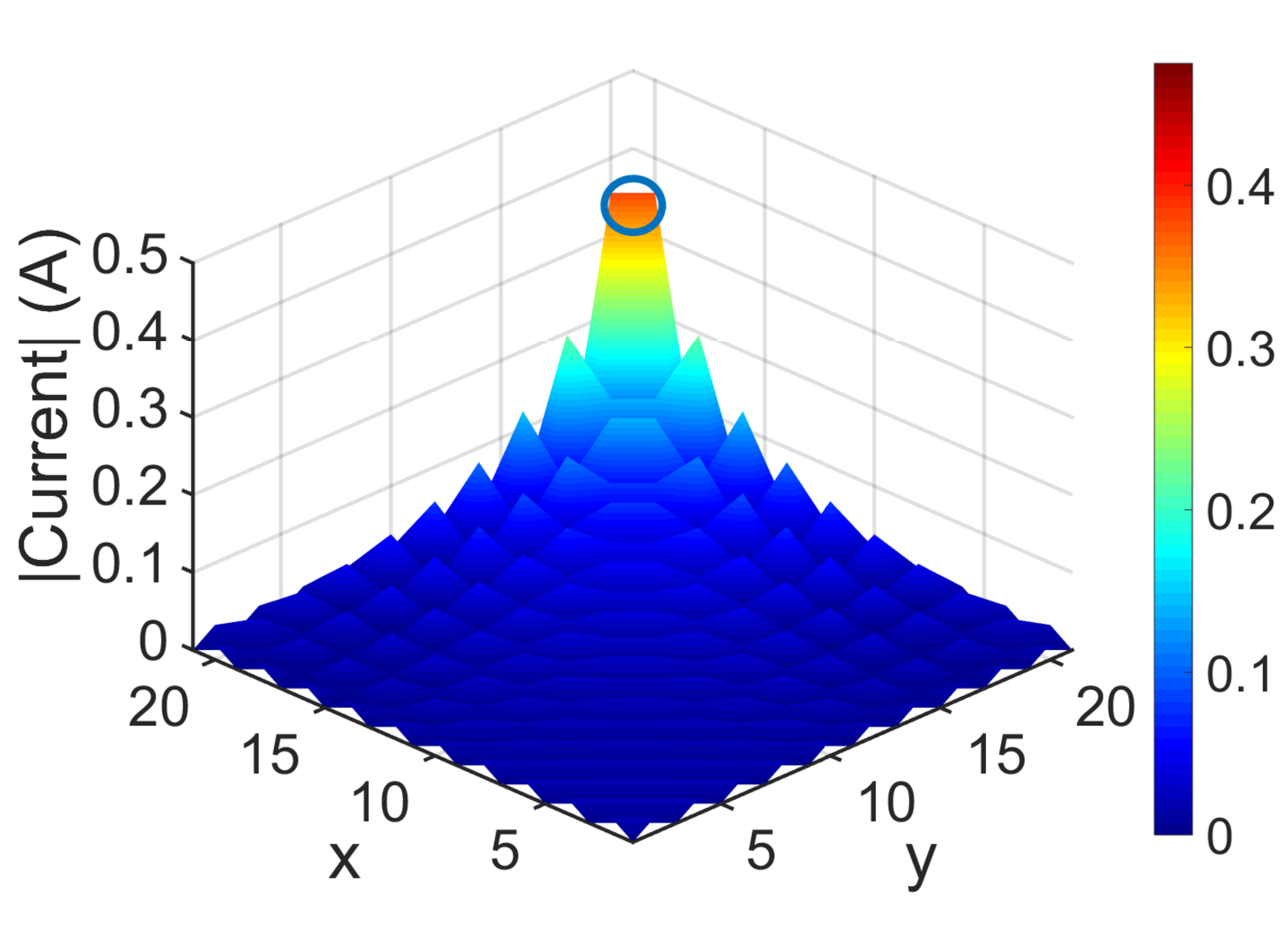}
			\label{fig:CurrentDist_10x10_Ideal_TLGrid_Gamma_Dirac}}}
	\caption{The voltage and current distributions across a single-point-fed $10\times10$ TL-grid array structure where the feeding point is highlighted by the blue circle.}
\end{figure}

The effect of the singularity is demonstrated by simulating a single-point-fed $10\times10$ array constructed from the symmetric TL-grid unit cell at the $\Gamma$-point frequency using Advanced Design System (ADS). The transmission lines used in the circuit model are assumed to be lossless and the 2D array is single-point-fed from the corner using a voltage source. The resulting voltage distribution across the 2D TL-grid array structure is presented in Fig.~\ref{fig:VoltageDist_10x10_Ideal_TLGrid_Gamma_Dirac} where the voltage is sampled at the individual unit cells' centers and boundaries. It is observed that the voltage distribution is uniform across the entire sampling locations within the 2D structure even though the grid is excited from a single point. It is crucial to highlight that the voltage distribution is uniform only at the individual unit cells' centers and boundaries and is nonuniform at other locations inside the unit cells.

On the other hand, the current distribution across the same array structure is depicted in Fig.~\ref{fig:CurrentDist_10x10_Ideal_TLGrid_Gamma_Dirac} with sampling at the individual unit cells boundaries. It is clear that the current distribution is not uniform. The current is maximum at the feeding location at the corner of the 2D structure and decays as the wave propagates away from the feeding point. The obvious discrepancy between the current and voltage profiles is a direct consequence of the singularity in the Z-matrix at the $\Gamma$-point where there is no longer a direct relation between the voltages and currents at the individual unit cell boundaries.

\section{2D Single-point Fed DLWA Design}

\begin{figure}    
	\centerline{
		\subfloat[Unit cell top view]{
			\includegraphics[width=0.21\textwidth]{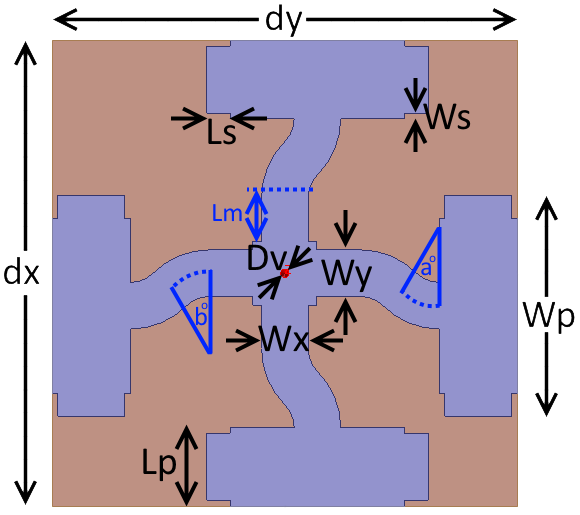}
			\label{fig:UnitCell_HFSS_TLGrid_Gamma_Patch_SCu}}
		\subfloat[Dispersion relation]{
			\includegraphics[width=0.27\textwidth]{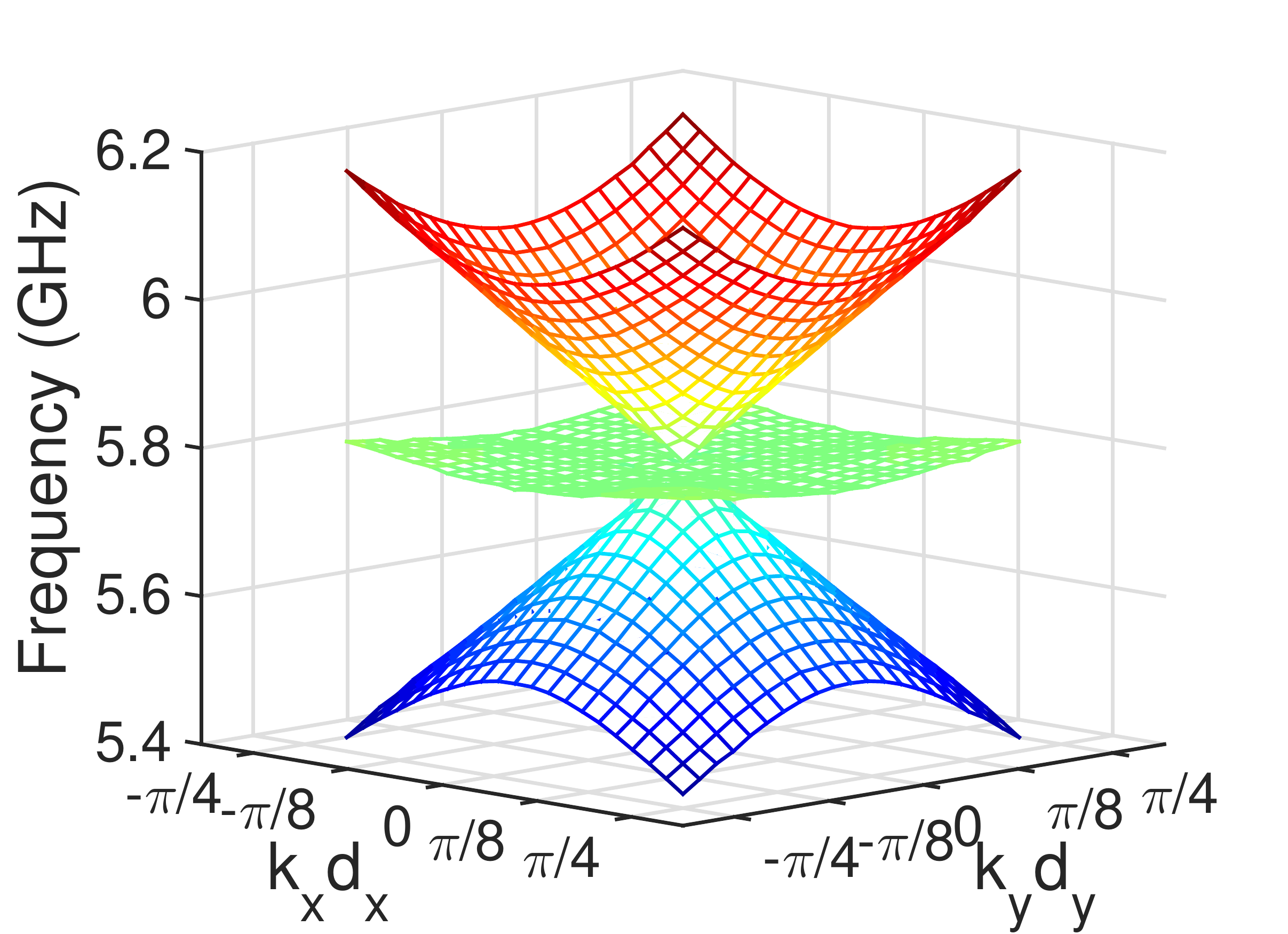}
			\label{fig:Dispersion_HFSS_TLGrid_Gamma_Patch_SCu_ClosedBandgap}}}
	\caption{Radiating TL-grid unit cell: (a) Top view of the unit cell (b) Full-wave simulated dispersion relation using HFSS. Unit cell dimensions: $d_x=d_y=40mm$, $d_{center~patch}=5.5mm$, $W_x=W_y=4mm$, $L_m=3.75mm$, $D_v=0.5mm$, $L_p=6.75mm$, $W_p=19mm$, $L_s=2mm$, $W_s=0.5mm$, $a^o=b^o=45^o$ and substrate thickness $=1.575mm$ on a Rogers RT/duriod 5880 substrate.}
\end{figure}

The symmetric TL-grid unit cell is modified into a radiating TL-grid unit cell by adding patches and curving the transmission lines as shown in Fig.~\ref{fig:UnitCell_HFSS_TLGrid_Gamma_Patch_SCu}. This opens the bandgap at the $\Gamma$-point which is subsequently closed by adding a metallic via and a square patch at the center of the unit cell. The simulated full-wave dispersion relation of the TL-grid LWA unit cell is depicted in Fig.~\ref{fig:Dispersion_HFSS_TLGrid_Gamma_Patch_SCu_ClosedBandgap} which confirms that the bandgap at the $\Gamma$-point is successfully closed. The closure of the bandgap means that the resulting unit cell is capable of generating a pencil beam at and around broadside. The bandgap is closed by achieving accidental degeneracy of the eigenmodes of interest at the $\Gamma$-point analogous to what has been proposed in \cite{DiracDegenracy}. It is also noticeable that the dispersion relation around the $\Gamma$-point varies linearly with frequency which means that the cones formed are Dirac cones and the corresponding antenna design is a Dirac leaky-wave antenna (DLWA) \cite{Moh_LWA}. It is important to mention that this DLWA TL-grid unit cell can be modeled by modifying the circuit model in Fig.~\ref{fig:UnitCell_Model_Ideal_TLGrid_Single} to account for the radiation of the unit cell and the effect of the patches and the metallic vias. However, it was preferred to close the bandgap by making the eigenmodes of interest at the $\Gamma$-point accidentally degenerate rather than using an equivalent circuit model.

\begin{figure} 
	\centerline{
		\hfill
		\subfloat{
			\includegraphics[width=0.16\textwidth,trim={{0.09\textwidth} 0 {0.12\textwidth} 0},clip]{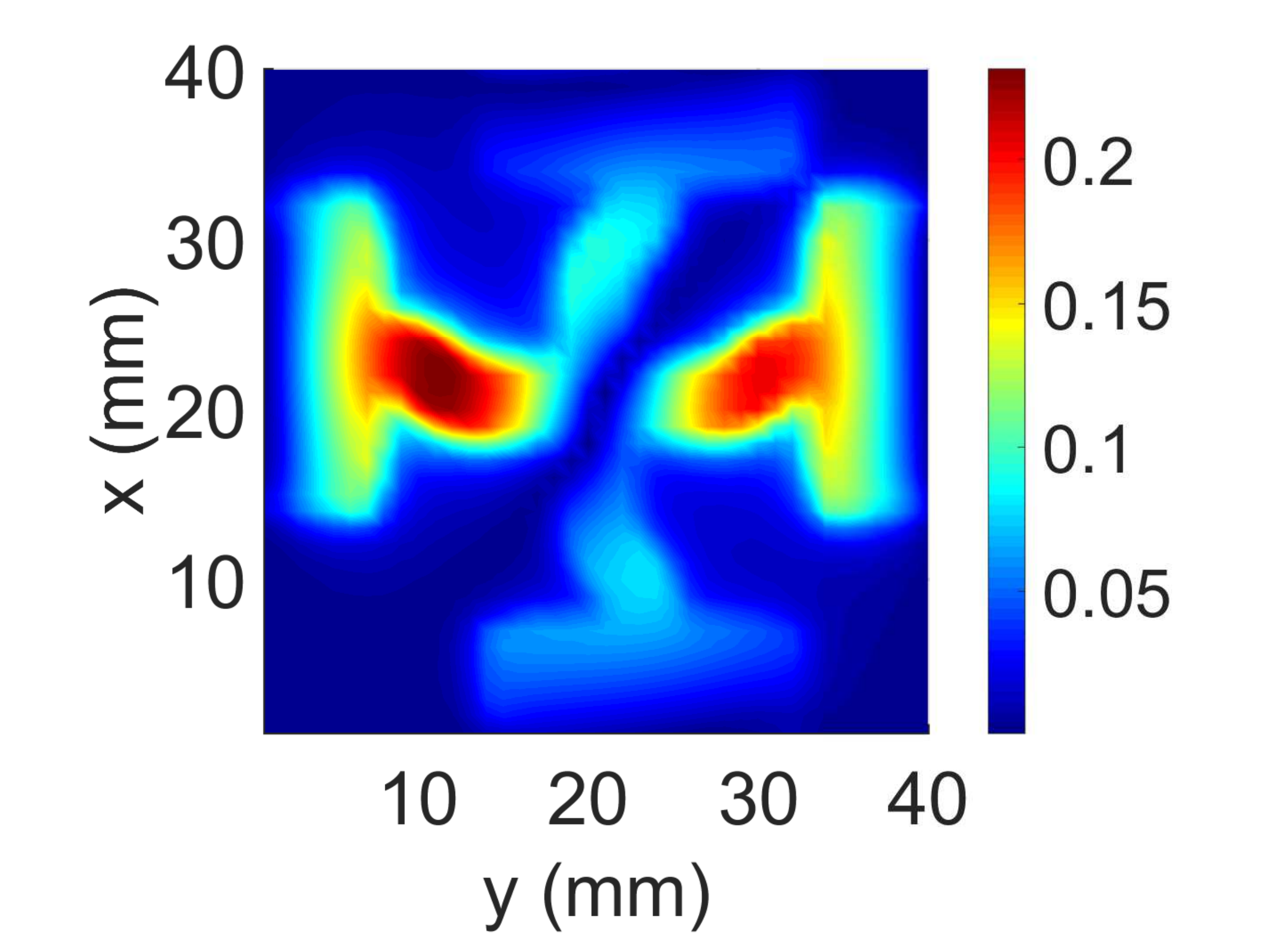}}
		\hfill
		\subfloat{
			\includegraphics[width=0.16\textwidth,trim={{0.09\textwidth} 0 {0.12\textwidth} 0},clip]{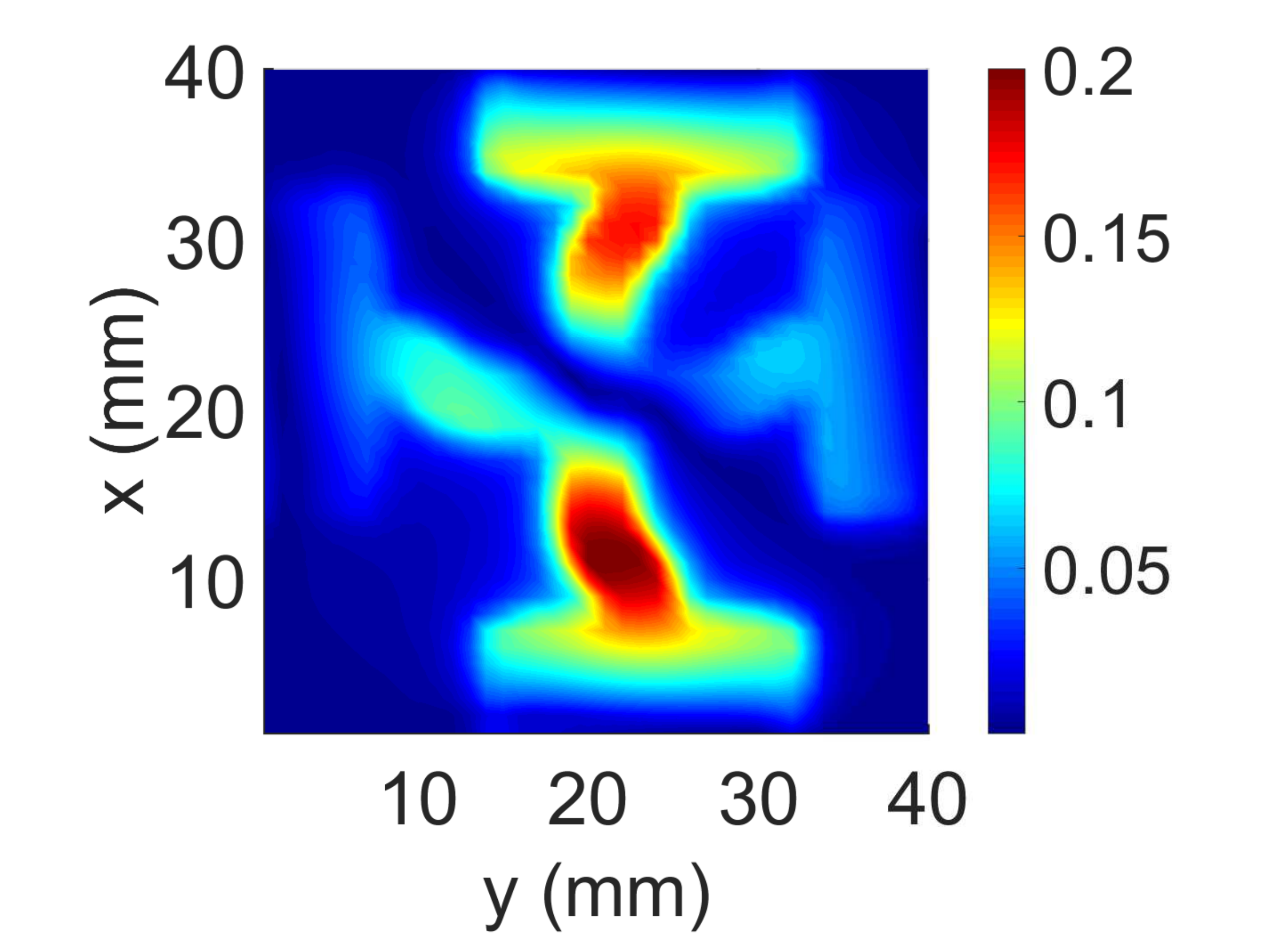}}
		\hfill
		\subfloat{
			\includegraphics[width=0.16\textwidth,trim={{0.09\textwidth} 0 {0.12\textwidth} 0},clip]{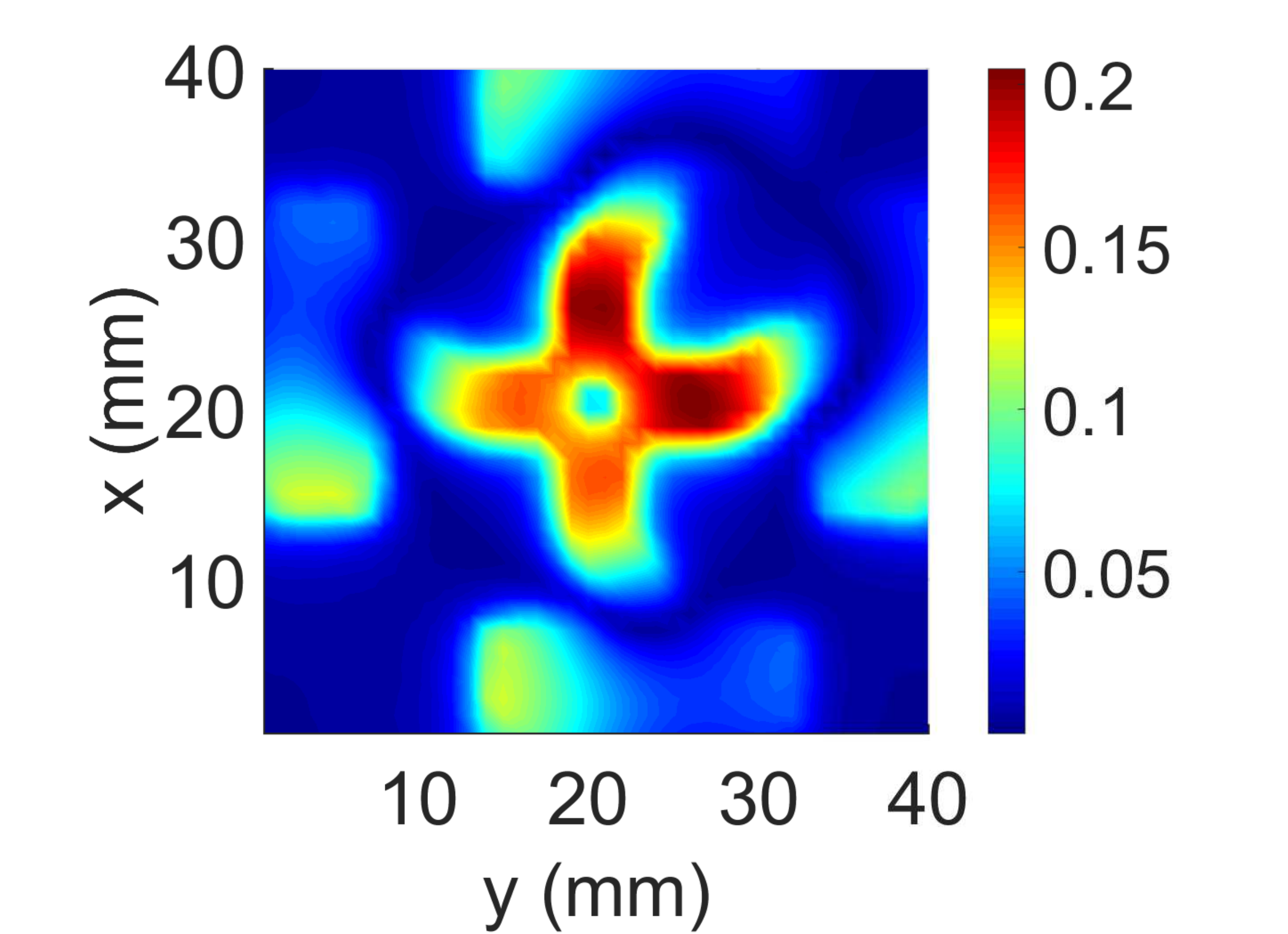}}
		\hfill}
	\caption{Full-wave eigenmode electric field magnitude [V/m] simulation results using HFSS at the $\Gamma$-point frequency ($5.77GHz$) for the unit cell shown in Fig.~\ref{fig:UnitCell_HFSS_TLGrid_Gamma_Patch_SCu}.}
	\label{fig:EH_HFSS_TLGrid_Gamma_Dirac_Patch_SCu}
\end{figure}

There is also a flat (deaf) band at the $\Gamma$-point frequency. Thus, at the $\Gamma$-point, there are 3 eigenmodes which were made accidentally degenerate and their field profiles are presented in Fig.~\ref{fig:EH_HFSS_TLGrid_Gamma_Dirac_Patch_SCu}. It is noticeable that the $1^{st}$ and $2^{nd}$ eigenmode fields resemble the electric fields present in a microstrip patch antenna at resonance. The patches are added at the unit cell boundaries because this is where the voltage distribution of a single-point-fed 2D array is uniform (Fig.~\ref{fig:VoltageDist_10x10_Ideal_TLGrid_Gamma_Dirac}). In addition, the transmission lines of the unit cell in Fig.~\ref{fig:UnitCell_HFSS_TLGrid_Gamma_Patch_SCu} are curved to enhance radiation from the $3^{rd}$ eigenmode. The 3 eigenmodes at the $\Gamma$-point are required to have balanced radiation levels to ensure a uniform gain when the beam is frequency scanned around broadside. In the proposed design, the radiation from the 3 eigenmodes is optimized to be as balanced as possible without reopening the bandgap in the dispersion relation.

\begin{figure}  
	\centering{
		\includegraphics[width=0.4\textwidth,trim={{0.07\textwidth} {0.22\textwidth} {0.07\textwidth} {0.21\textwidth}},clip]{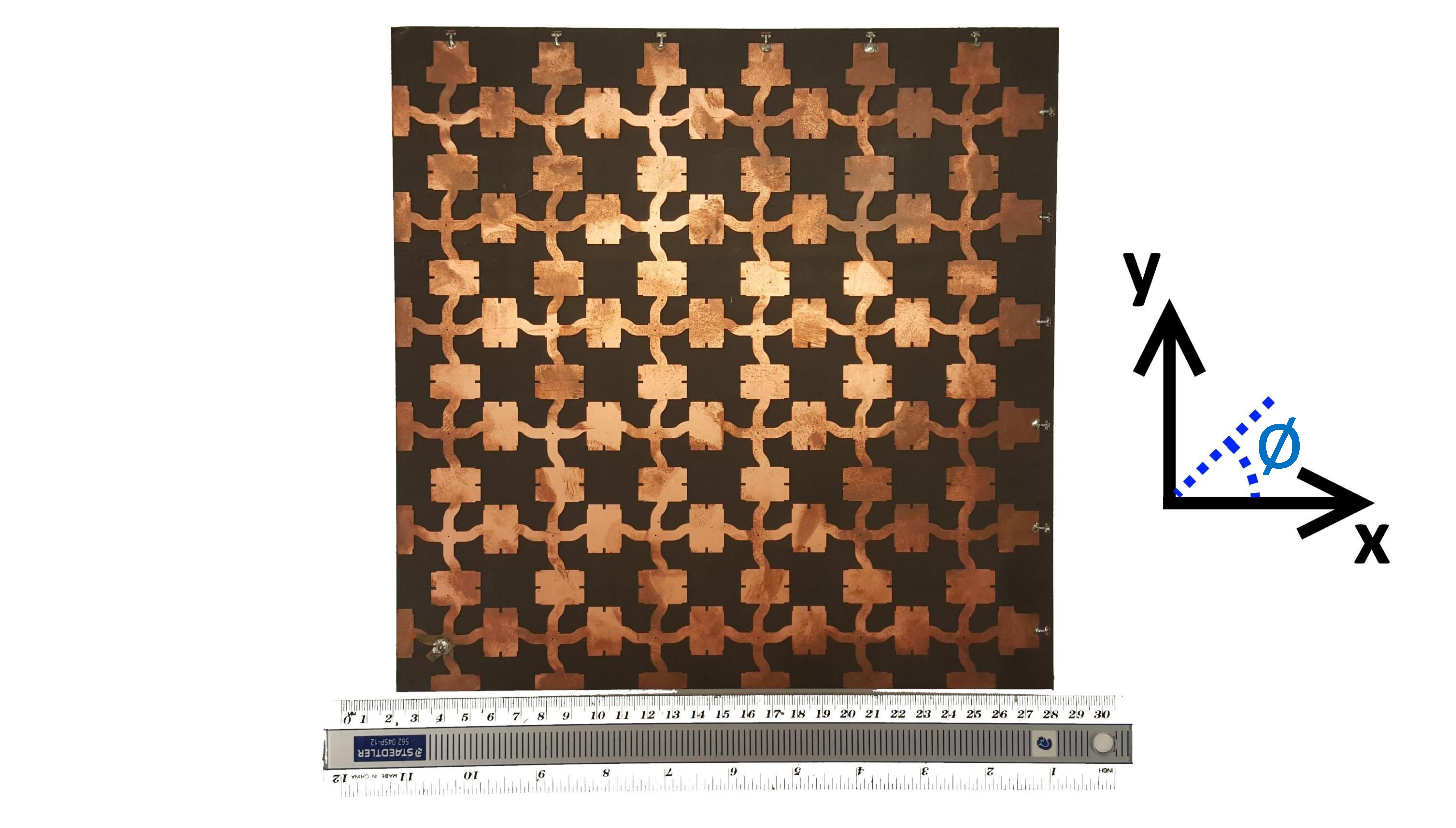}}
	\caption{Top view photo of the fabricated 2D DLWA (unit cell in Fig.~\ref{fig:UnitCell_HFSS_TLGrid_Gamma_Patch_SCu}) with dimensions $254.75\times254.75\times1.575~mm$.}
	\label{fig:2D_6x6_Meas_TLGrid_Gamma_Patch_SCu}
\end{figure}

The optimized radiating unit cell is arranged in a $6\times6$ array and the fabricated structure is shown in Fig.~\ref{fig:2D_6x6_Meas_TLGrid_Gamma_Patch_SCu}. This 2D structure is excited from the bottom-left corner using a single coaxial feed coming from below the 2D DLWA. 
An Open Circuit (OC) parallel stub is used at the input to match the design to the $50\Omega$ coaxial line. As for the terminations from the top and right sides of the design, the 2D DLWA design is effectively terminated to the Bloch impedance value. 

The 2D DLWA was left open circuited from the bottom and left sides (surrounding the coaxial feed) to avoid losing power directly to the terminations without ever going to the DLWA. Thus, the realized gain by leaving the bottom and left sides OC is increased as opposed to using Bloch impedance terminations (resistors) without severely perturbing proper TL-grid operation. 

\begin{figure}[t] 
	\centerline{
		\subfloat[$S_{11}$ response]{		
			\includegraphics[width=0.24\textwidth,trim={{0.015\textwidth} {0} {0.08\textwidth} {0.05\textwidth}},clip]{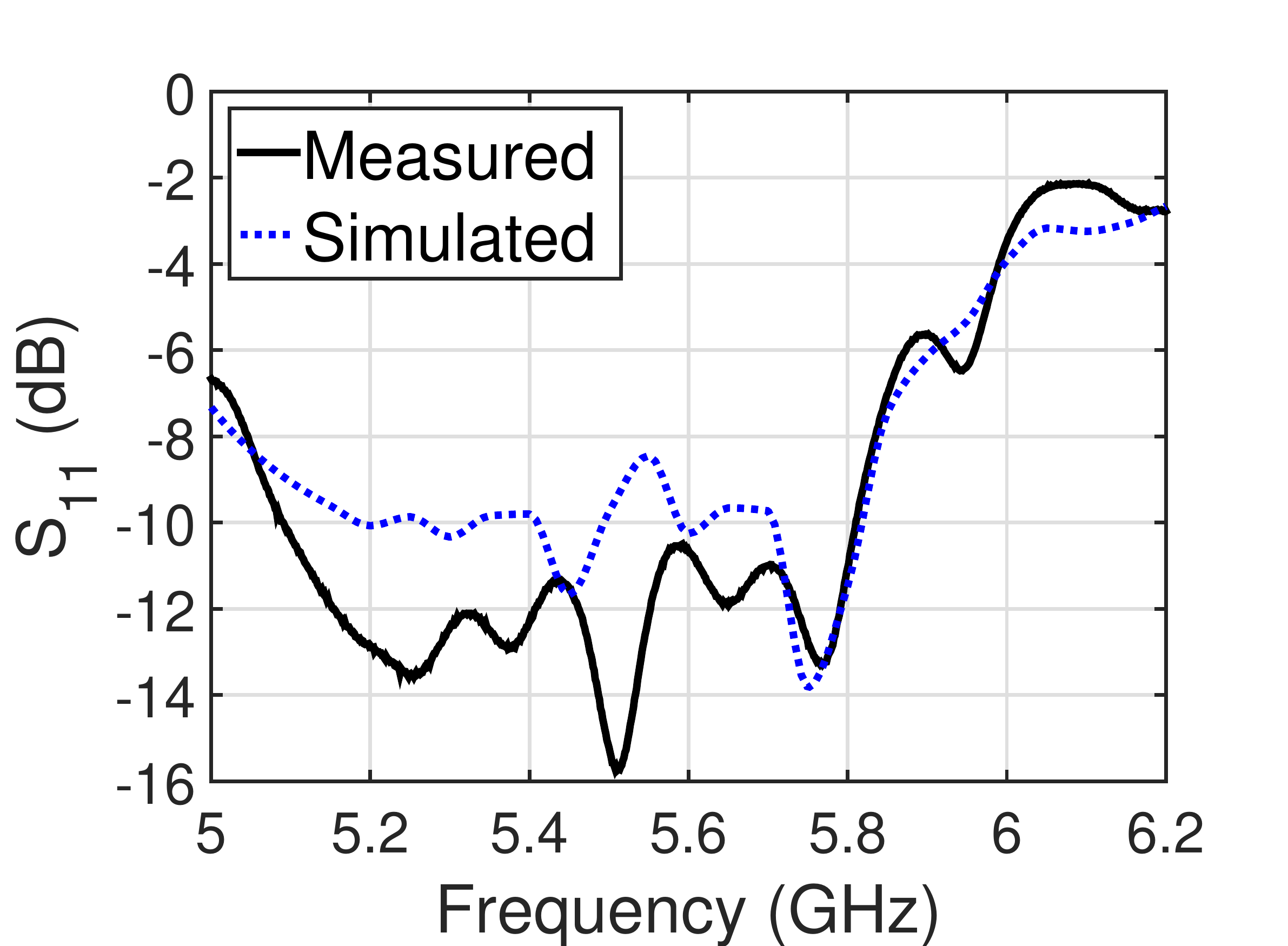}
			\label{fig:SPar_6x6_Meas_TLGrid_Gamma_Patch_SCu_All}}
		\subfloat[Radiation Efficiency]{
			\includegraphics[width=0.24\textwidth,trim={{0.02\textwidth} {0} {0.08\textwidth} {0.05\textwidth}},clip]{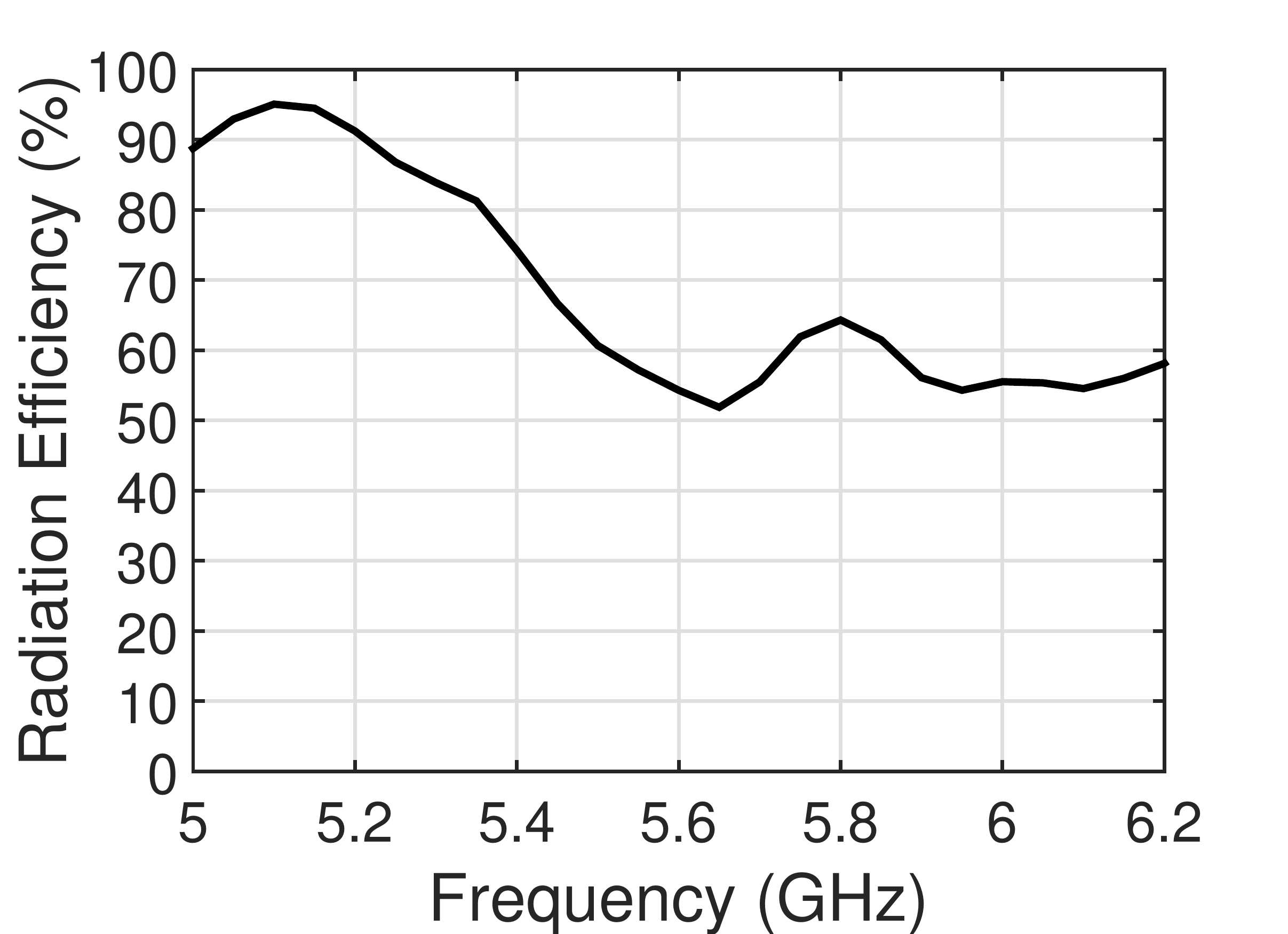}
			\label{fig:RadEfficiency_6x6_HFSS_TLGrid_Gamma_Patch_SCu_All}}}
	\caption{(a) Measured and simulated $S_{11}$ response (b) Simulated radiation efficiency ($\%$).}
\end{figure}

The full-wave simulated and measured $S_{11}$ are plotted in Fig.~\ref{fig:SPar_6x6_Meas_TLGrid_Gamma_Patch_SCu_All}. The simulated and measured $S_{11}$ are below $8dB$ and $10dB$, respectively, throughout the frequency range of interest. On the other hand, Fig.~\ref{fig:RadEfficiency_6x6_HFSS_TLGrid_Gamma_Patch_SCu_All} depicts the simulated radiation efficiency of this 2D DLWA and its behavior with frequency. The calculated radiation efficiency is between $60\%$-$90\%$ in the frequency range of interest and is $63.5\%$ at broadside where the remaining unradiated power is absorbed in the terminations. 

\begin{figure}   
	\centerline{
		\subfloat[E-plane ($\phi=45^o$)]{
			\includegraphics[width=0.24\textwidth,trim={{0.02\textwidth} {0} {0.08\textwidth} {0.05\textwidth}},clip]{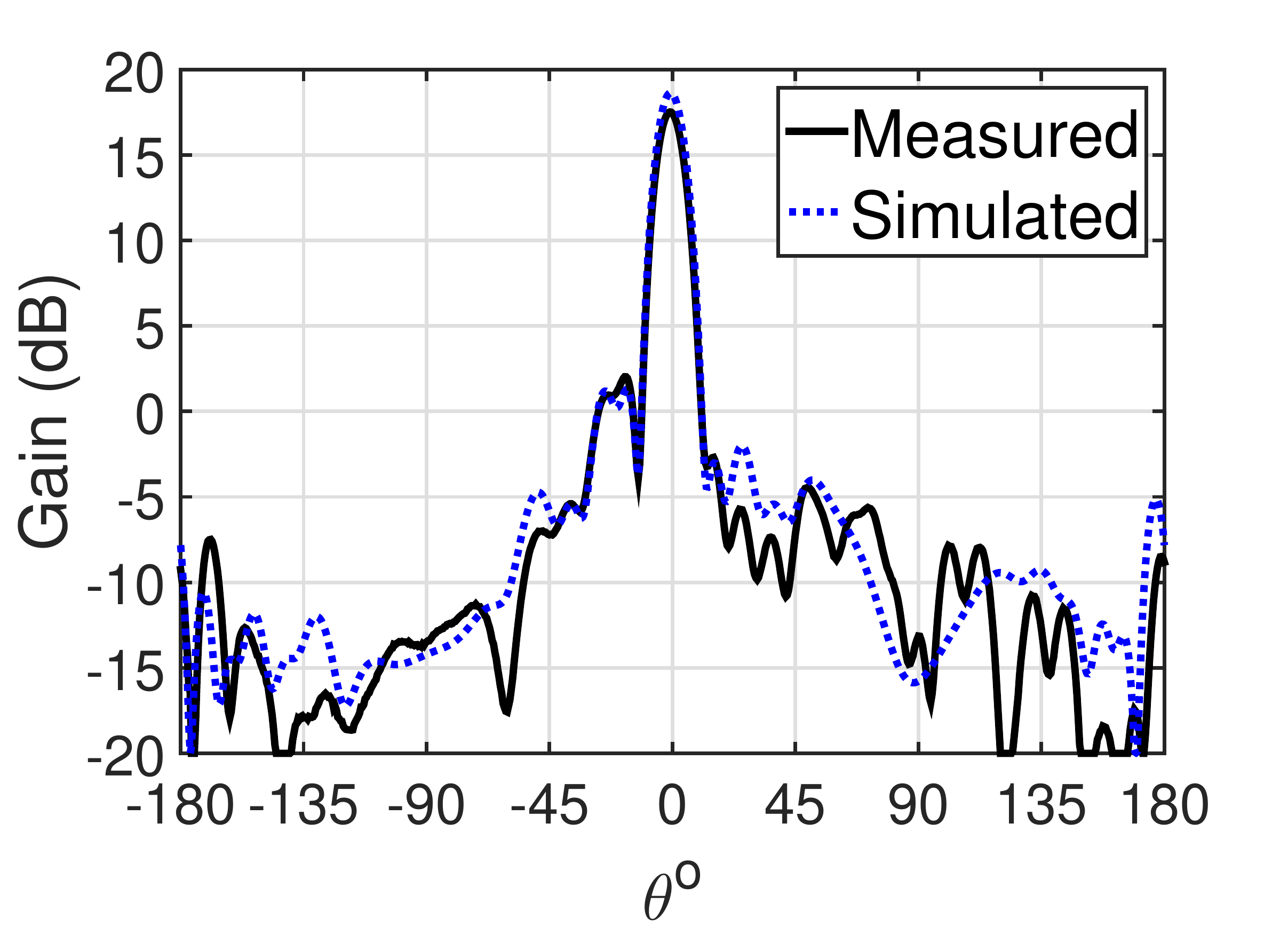}}
		\subfloat[H-plane ($\phi=135^o$)]{
			\includegraphics[width=0.24\textwidth,trim={{0.02\textwidth} {0} {0.08\textwidth} {0.05\textwidth}},clip]{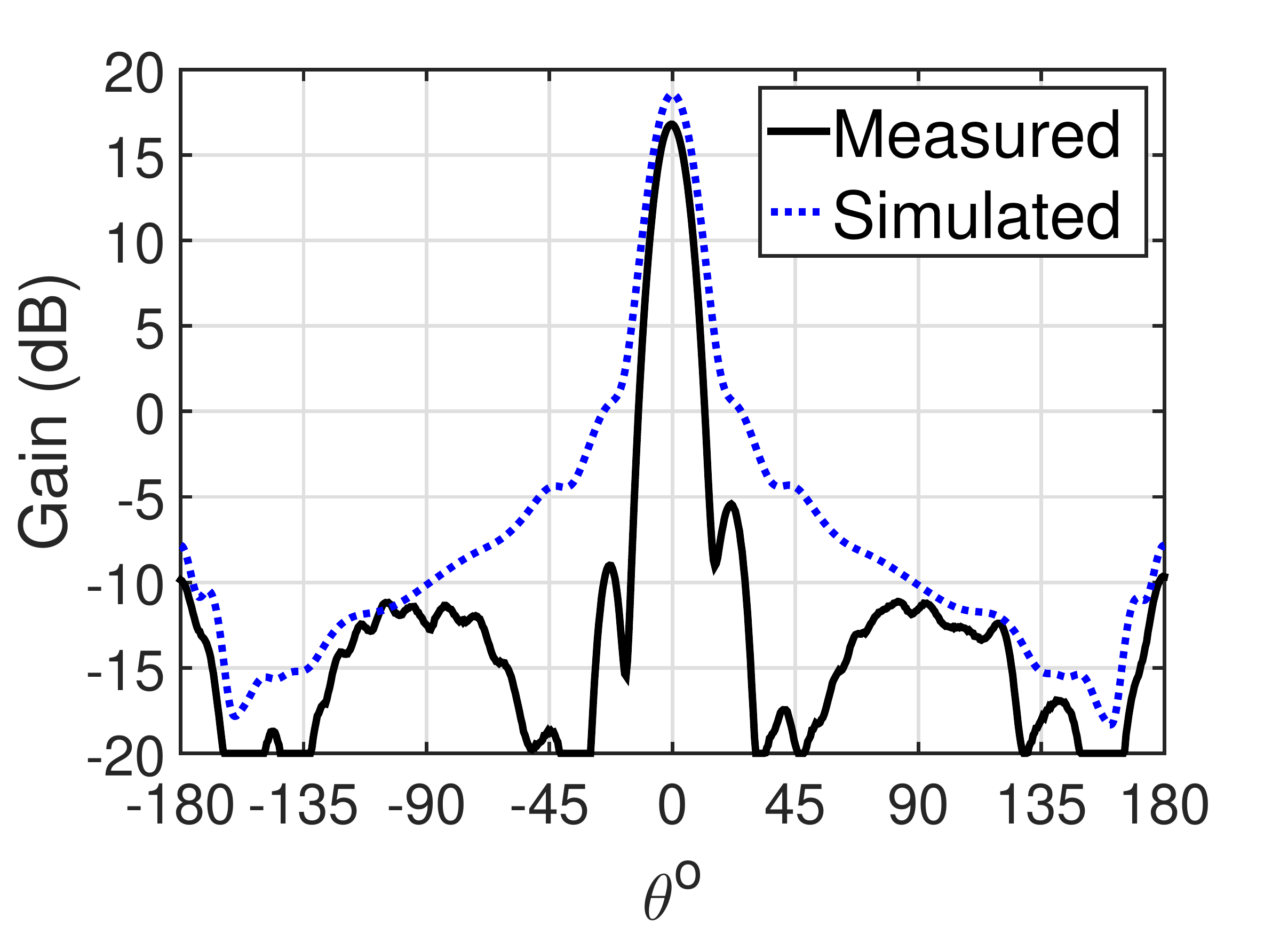}}}
	\caption{Measured gain and simulated realized gain in $40dB$-scale at broadside ($5.75GHz$).}
	\label{fig:RGain_6x6_MeasHFSS_TLGrid_Gamma_Patch_SCu_Dirac_AllPlanes}
\end{figure}

At the $\Gamma$-point ($5.75GHz$), the measured gain is compared against the full-wave simulated realized gain in Fig.~\ref{fig:RGain_6x6_MeasHFSS_TLGrid_Gamma_Patch_SCu_Dirac_AllPlanes}. A measured gain of $17.5dB$ is achieved compared to a simulated one of $18.6dB$. The two gain plots also show the same trend of the sidelobes and the overall beam shape. 

\begin{figure}[t]  
	\centerline{
		\subfloat[$5.0GHz$]{
			\includegraphics[width=0.165\textwidth,trim={{0.25\textwidth} {0.09\textwidth} {0.22\textwidth} {0.06\textwidth}},clip]{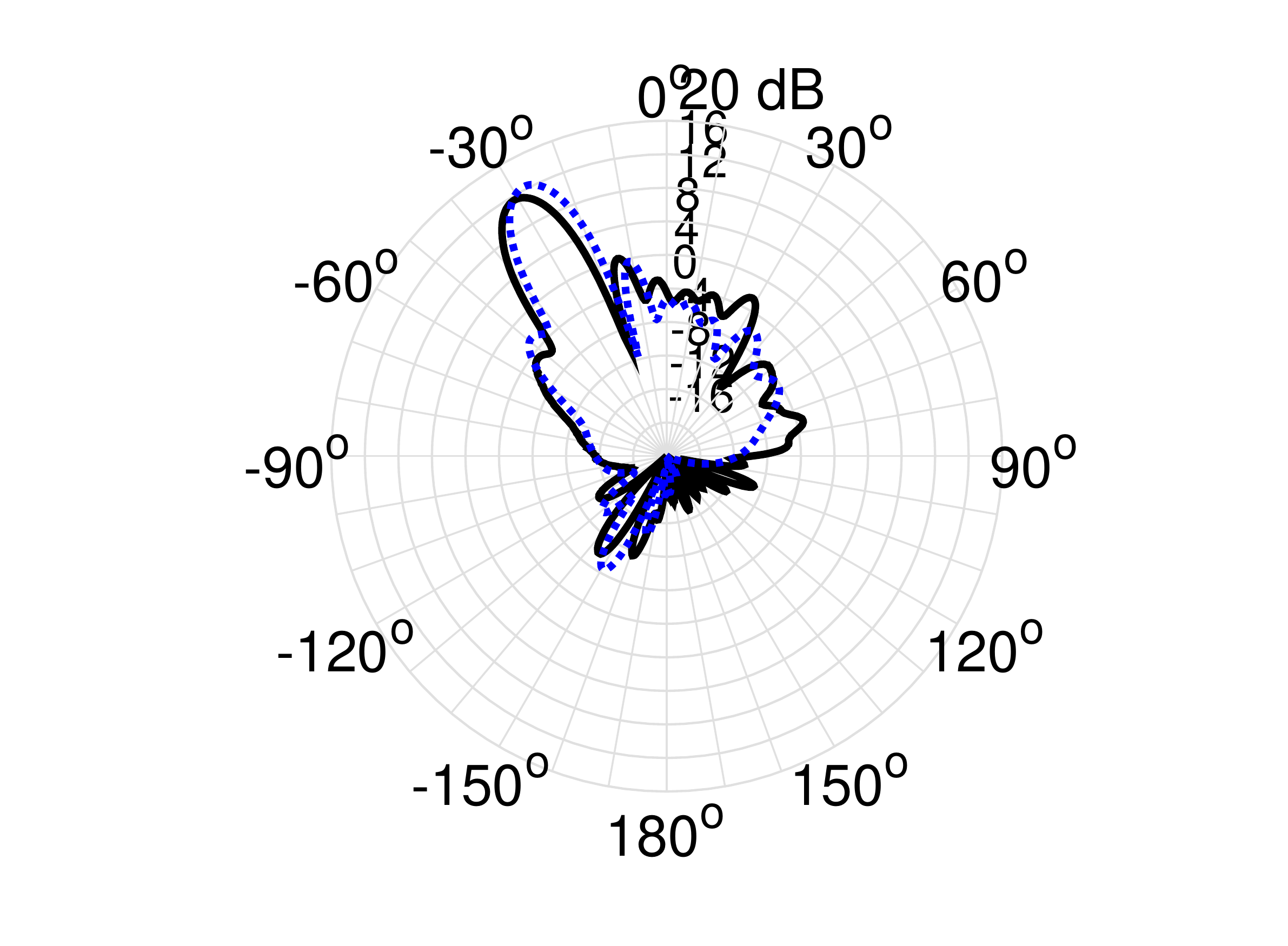}}
		\subfloat[$5.2GHz$]{
			\includegraphics[width=0.165\textwidth,trim={{0.25\textwidth} {0.09\textwidth} {0.22\textwidth} {0.06\textwidth}},clip]{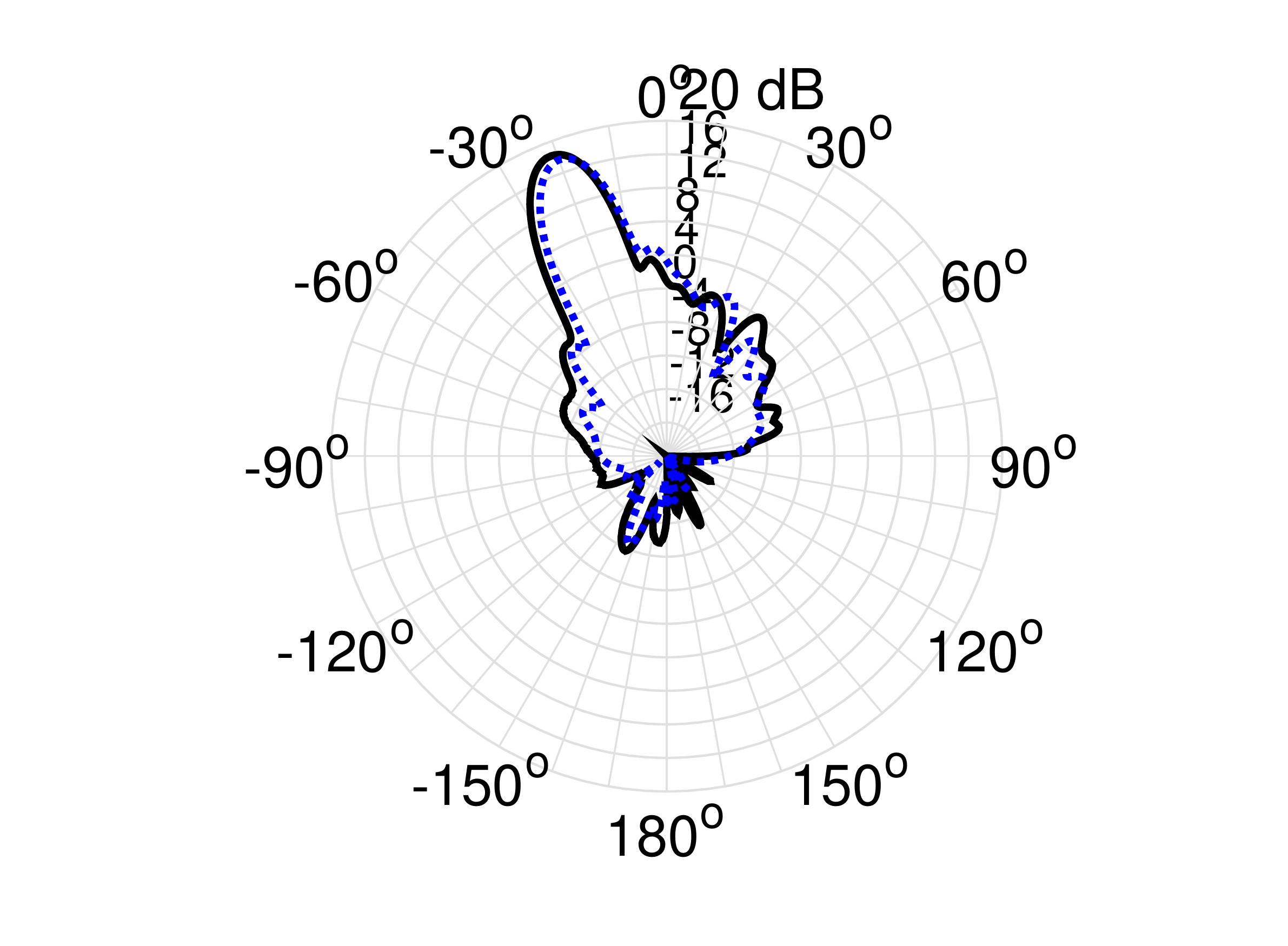}}
		\subfloat[$5.4GHz$]{
			\includegraphics[width=0.165\textwidth,trim={{0.25\textwidth} {0.09\textwidth} {0.22\textwidth} {0.06\textwidth}},clip]{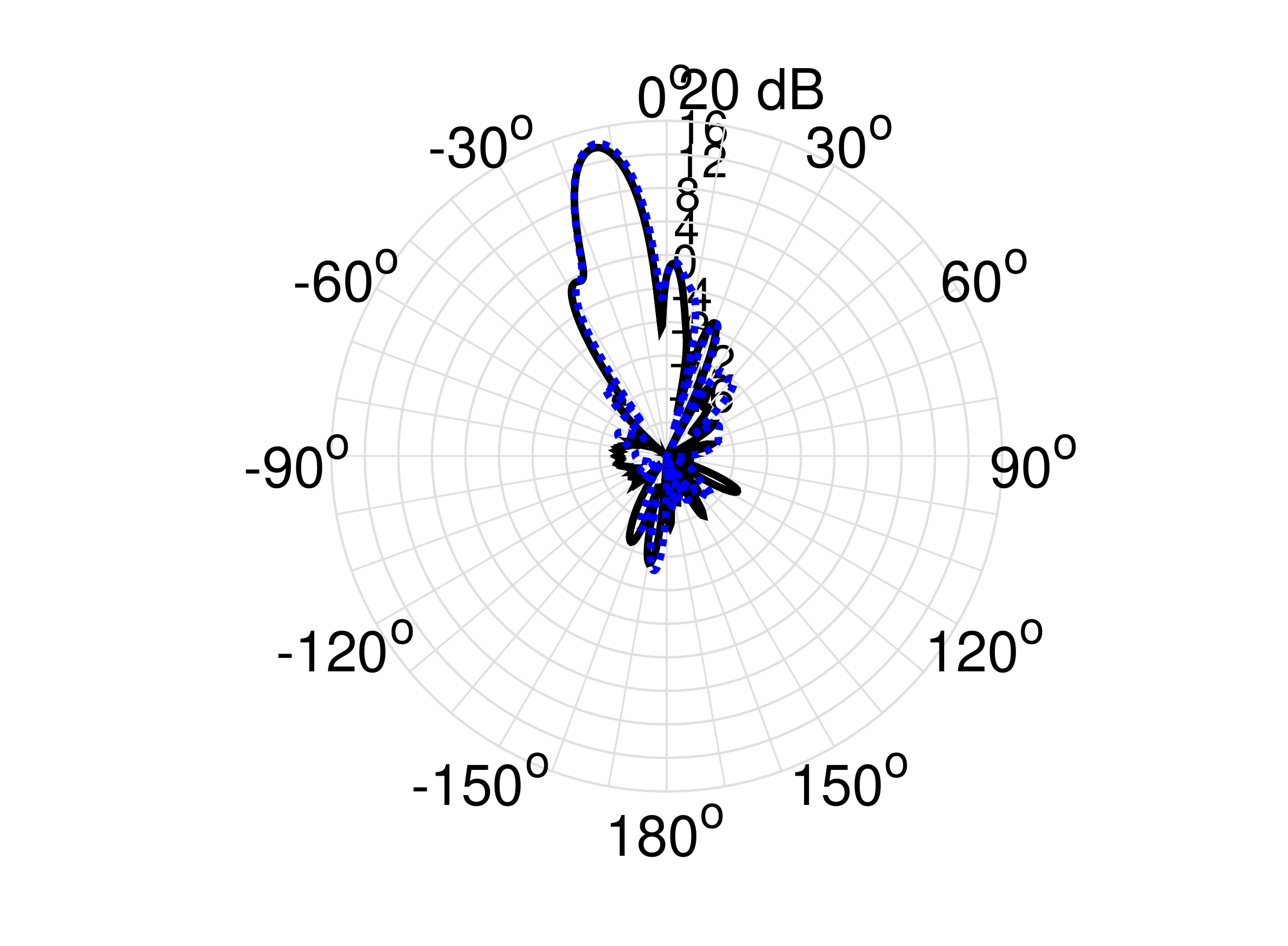}}}
	\centerline{
		\subfloat[$5.6GHz$]{
			\includegraphics[width=0.165\textwidth,trim={{0.25\textwidth} {0.09\textwidth} {0.22\textwidth} {0.06\textwidth}},clip]{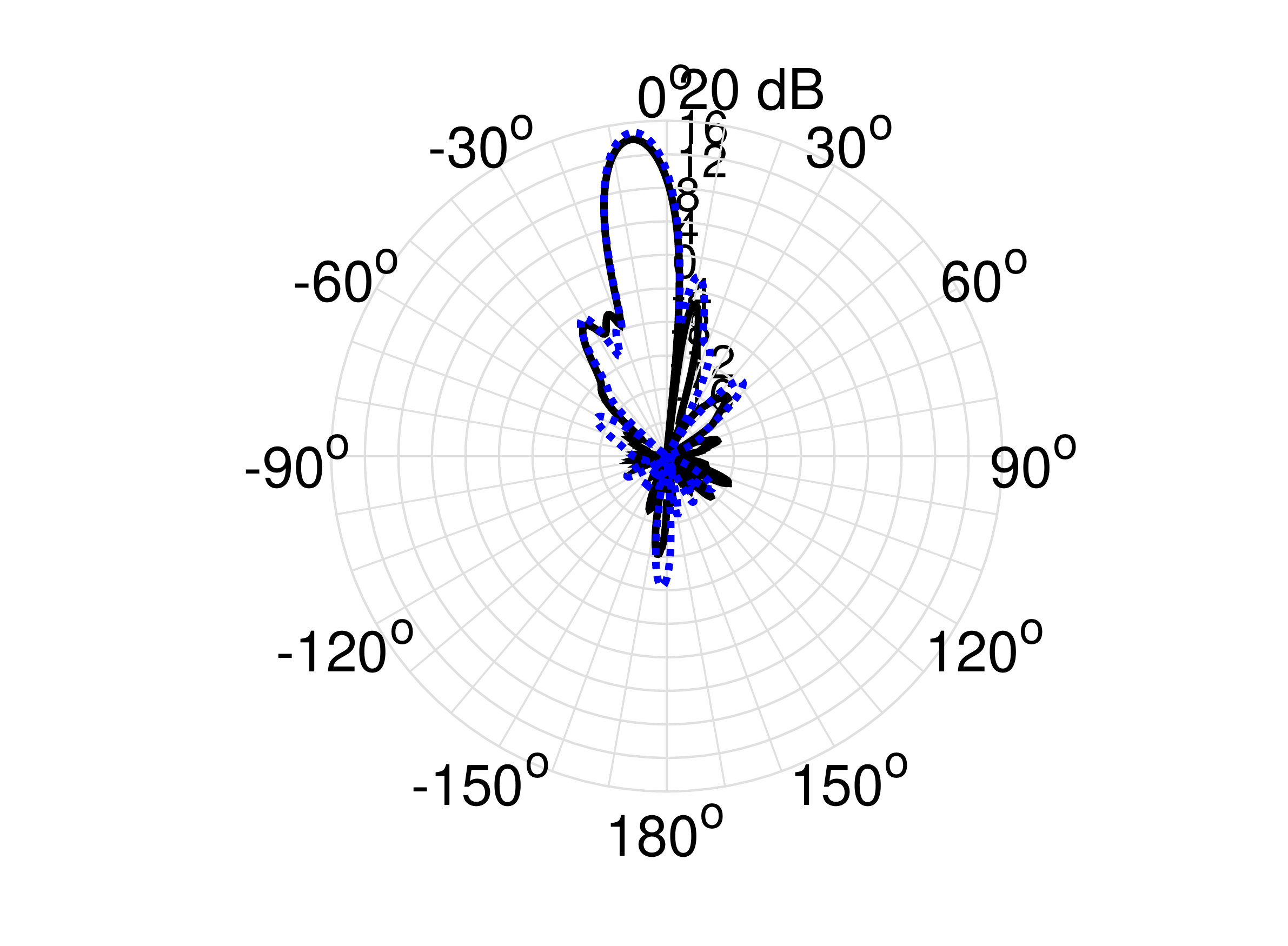}}
		\subfloat[$5.9GHz$]{
			\includegraphics[width=0.165\textwidth,trim={{0.25\textwidth} {0.09\textwidth} {0.22\textwidth} {0.06\textwidth}},clip]{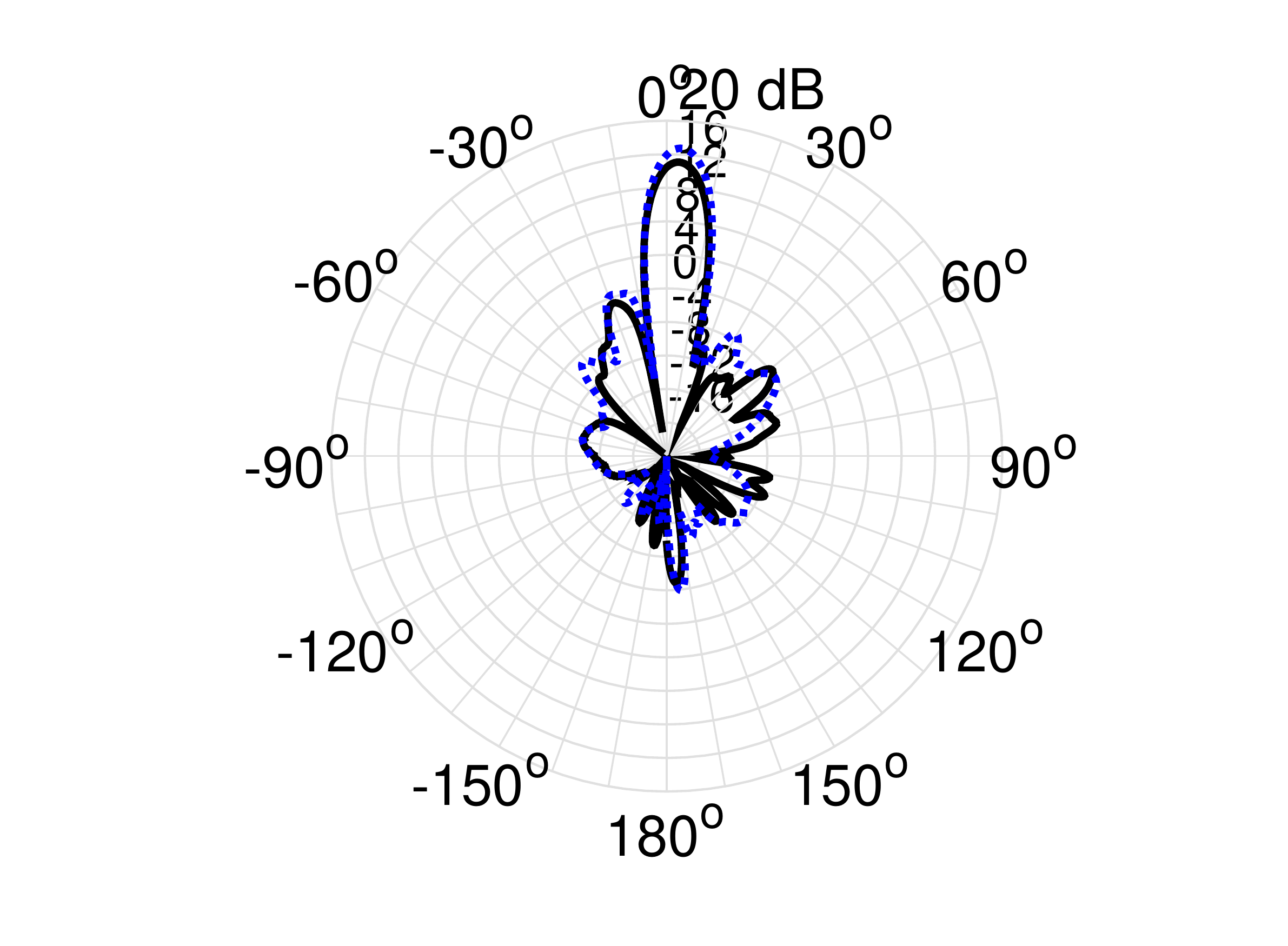}}
		\subfloat[$6.2GHz$]{
			\includegraphics[width=0.165\textwidth,trim={{0.25\textwidth} {0.09\textwidth} {0.22\textwidth} {0.06\textwidth}},clip]{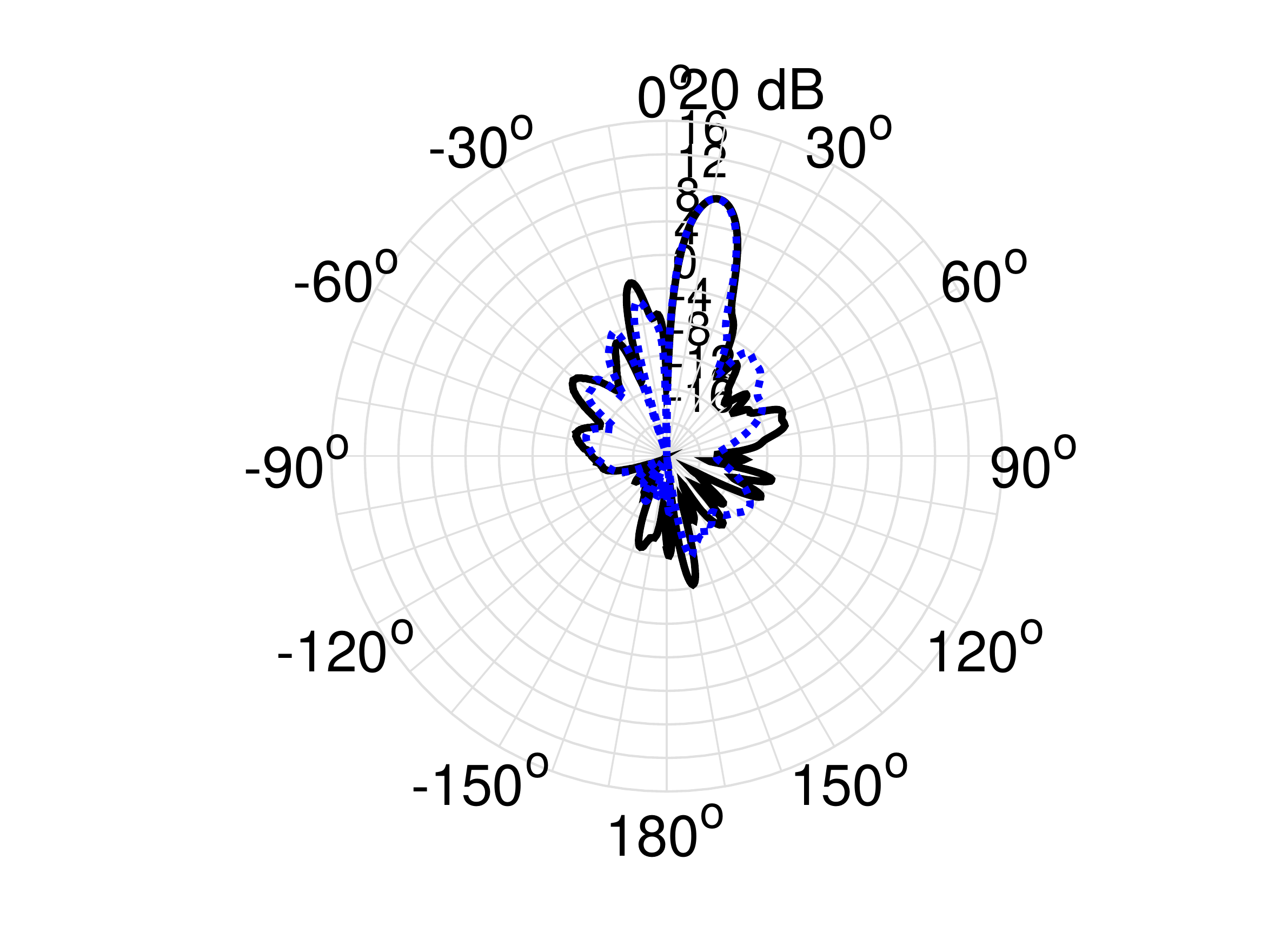}}}
	\caption{E-plane ($\phi=45^o$) measured gain (solid black) and simulated realized gain (dotted blue) in $40dB$-scale.}
	\label{fig:RGain_6x6_MeasHFSS_TLGrid_Gamma_Patch_SCu_AllFreq_EPlane}
\end{figure}

The actual measured gain and simulated realized gain patterns of this 2D DLWA are plotted in Fig.~\ref{fig:RGain_6x6_MeasHFSS_TLGrid_Gamma_Patch_SCu_AllFreq_EPlane} in $40dB$-scale for an E-Plane cut ($\phi=45^o$) where the beam is frequency scanned from $5$-$6.2GHz$ ($\theta_{max}=-31^o$ to $11.5^o$). It is clear that this 2D DLWA is capable of forward and backward radiation with a uniform gain level and without breaking at broadside which is an indication of successful closing of the bandgap. The measured realized gain remains fairly uniform as the beam is frequency scanned in the backward radiation direction. However, as the beam is scanned in the forward direction, there is a decrease in gain because of the emergence of grating lobes at higher frequencies. The measured gain plots in this E-Plane cut are shifted by $\theta=+3.5^o$ to partially account for angular misalignments during testing.

Furthermore, a comparison between the measured gain and the simulated realized gain is presented in Table~\ref{tbl:RGainComparison_6x6_MeasHFSS_TLGrid_Gamma_Patch_SCu_All_EPlanes} for the same frequency range. Indeed both have the same level of gain and performance with some discrepancies that can be attributed to misalignments during testing.
\begin{table}
	\renewcommand{\arraystretch}{1.05}
	\caption{Actual and simulated gains for the E-Plane cut.}
	\label{tbl:RGainComparison_6x6_MeasHFSS_TLGrid_Gamma_Patch_SCu_All_EPlanes}
	\centering
	\begin{tabular}{|c | c | c |}
		\hline
		Frequency (GHz) 	& Simulated Realized Gain	   & Measured Gain     \\
		\hline
		$5.0GHz$			&		  $16.4dB$  		   &	 $15.5dB$	   \\
		$5.2GHz$			&		  $17.6dB$			   &	 $18.4dB$	   \\
		$5.4GHz$			&		  $18.2dB$  		   &	 $17.7dB$	   \\
		$5.6GHz$			&		  $18.8dB$			   &	 $18.0dB$	   \\
		$5.75GHz$           &         $18.6dB$             &     $17.5dB$      \\
		$5.9GHz$			&		  $16.7dB$  		   &	 $15.1dB$	   \\
		$6.2GHz$			&		  $11.3dB$			   &	 $11.3dB$	   \\
		\hline
	\end{tabular}	
\end{table}
In contrast to the design presented in \cite{CalozConical} which generates a conical beam, the proposed DLWA TL-grid design is capable of generating a highly directive pencil beam at both broadside and tilted angles because of the closed bandgap in the dispersion relation, the corner feed used and other important factors such as the larger unit cell size of the TL-grid unit cell when compared to typical metamaterial unit cell sizes and the open circuit terminations adopted around the coaxial feed. The generated pencil beam is frequency scanned in the E-plane of the 2D DLWA with an acceptable level of matching and because of the established Dirac cones the scanning of the beam is linear with frequency. On the other hand, the proposed TL-grid DLWA design is analogous to the photonic crystal DLWA design presented in \cite{Moh_LWA} where both designs employ the accidental degeneracy of the eigenmodes of interest at the $\Gamma$-point to establish the Dirac cones and have continuous beam steering through broadside. However, the DLWA design presented in \cite{Moh_LWA} is 1D, dielectric based and generates a fan beam whereas the proposed TL-grid DLWA design here is 2D, transmission-line based and generates a pencil beam.

\section{Conclusion}
In this paper, we have used S-parameter analysis to show how to design a TL grid with a uniform voltage distribution (magnitude and phase) at its nodes. It has also been shown by circuit simulations that this uniform voltage distribution is maintained at the $\Gamma$-point even for a single-point-fed 2D array. The corresponding TL-grid unit cell has been modified into a radiating unit cell and it has been shown that it is possible to close the bandgap in the dispersion relation by establishing accidental degeneracy of the eigenmodes of interest at the $\Gamma$-point. The dispersion relation of the radiating unit cell has also been shown to exhibit a closed bandgap with cones that are linearly varying with frequency which are the characteristics of a Dirac dispersion. Based on this, a TL-grid 2D Dirac LWA design was presented. The antenna elements were added at the unit cell boundaries to ensure the uniformity of the individual antenna elements' excitation level even when fed from a single feeding point. The fabricated 2D DLWA is single-point-fed from the corner, thus, does not require a complex feeding network. In addition, the proposed design can achieve continuous beam scanning from backward to forward radiation directions with a linear variation with frequency and without breaking at broadside. This experimentally demonstrates the possibility of using the TL-grids as single-point-fed 2D DLWAs where the underlying grid achieves the required amplitude and phase excitation of the antenna elements.

%\section*{Acknowledgment}
%For the Summary paper submission only, no acknowledgements are allowed. 

%% This adds a line for the Bibliography in the Table of Contents.
\addcontentsline{toc}{chapter}{Bibliography}
%% *** Set the bibliography style. ***
%% (change according to your preference/requirements)
\bibliographystyle{ieeetran}
%% *** Set the bibliography file. ***
%% ("thesis.bib" by default; change as needed)
\bibliography{bibliography}
\smallskip
\end{document}